\documentclass[aps,prd,preprint,groupedaddress]{revtex4}

\usepackage{epsf}
\usepackage{longtable}
\usepackage{amsfonts}
\usepackage{graphicx}

\usepackage{epsf}

\def\Hz{{\rm Hz}}
\def\km{{\rm km}}
\def\AU{{\rm AU}}

\def\obs{{\rm obs}}
\def\kms{{\rm km/s}}
\def\matter{{\rm matter}}

\def\kHz{{\rm kHz}}
\def\mas{{\rm mas}}
\def\years{{\rm years}}
\def\const{{\rm const}}
\def\delay{{\rm delay}}
\def\time{{\rm time}}

\def\obs{{\rm obs}}
\def\max{{\rm max}}

\def\Mean{{\rm Mean}}

\newcommand{\simgt}{\lower.5ex\hbox{$\; \buildrel > \over \sim \;$}}
\newcommand{\simlt}{\lower.5ex\hbox{$\; \buildrel < \over \sim \;$}}

\newcommand{\baredth}{\;\overline{\raise1.0pt\hbox{$'$}\hskip-6pt
\partial}\;}
\newcommand{\edth}{\;\raise1.0pt\hbox{$'$}\hskip-6pt\partial\;}

\newcommand{\gsim}{\lower.5ex\hbox{$\; \buildrel > \over \sim \;$}}
\newcommand{\lsim}{\lower.5ex\hbox{$\; \buildrel < \over \sim \;$}}
\def\aap  {A\&A}

\def\apj  {Astrophys. J.}
\def\apjl {Astrophys. J. Lett.}

\def\nat  {Nature}
\def\prd  {Phys.~Rev.~D.}
\def\ouraffiliation {{Astronomical Institute, Graduate School of Science, 
Tohoku University, Sendai 980-8578, Japan}}

\begin{document}

\title{A method to measure a relative transverse velocity of 
source-lens-observer system using gravitational 
lensing of gravitational waves}

\author{Yousuke Itoh}
\email{yousuke@astr.tohoku.ac.jp}
\affiliation{\ouraffiliation}
\author{Toshifumi Futamase}
\affiliation{\ouraffiliation}
\author{Makoto Hattori}
\affiliation{\ouraffiliation}

\begin{abstract}
Gravitational waves propagate 
along  null geodesics like light rays in the geometrical optics approximation, 
and they may have a chance to 
suffer from gravitational lensing by intervening objects, as 
is the case for electromagnetic waves.  
Long wavelength of gravitational waves and compactness of 
possible sources may enable us to extract information in the interference among the lensed images.  We point out that 
the interference term contains information of relative 
transverse velocity of the source-lens-observer system, which 
may be obtained by possible future space-borne gravitational wave detectors 
such as BBO/DECIGO.
\end{abstract} 

\maketitle

\section{Introduction}

Direct observations of gravitational waves 
from astrophysical sources would open a new window to the universe, and 
a new era of 
astronomy. 
It is now well known that there are many gravitational lens systems of
various sorts in the electromagnetic astronomy. 
This must also happens in gravitational wave astronomy 
in the future. 
One difference between electromagnetic 
waves and gravitational waves are their wavelength.  
The wavelength of gravitational waves which the currently working 
and the planned gravitational wave observatories may detect 
are typically much larger than the electromagnetic waves in the
current astronomy. Therefore once they suffer from gravitational
lensing, we have a chance to observe interference. 
To be more quantitatively, 
a condition may be met for interference to be
seen. In the electromagnetic astronomy, a wave source is
typically extended and it consists of pieces of incoherent 
emitters. Then, to observe interference, the typical linear dimension 
of the source $l$ should 
be smaller than or comparable to the wavelength $\lambda$ 
divided by the angular separation among the multiple images $\theta$ (in radians), or  
$l \lsim \kappa \lambda/\theta \equiv \kappa l_0$ with $\kappa$ of order unity in 
a typical gravitational lensing occurring in cosmology.   
Otherwise incoherent waves from different parts of the source distract the
interference.  Observing at, say, 
$\lambda = 1\mu$m 
and assuming 
$\theta = 10'' \simeq 5\times  10^{-5}$ radians, 
$l_0 \simeq 2$ cm.  This is one reason why we usually 
do not see   
interference in the electromagnetic 
astronomy. One exception where wave effects become important 
is scintillation for radio sources due to ionized interstellar/interplanetary gas 
(See, e.g., Chap. 13.4 of \cite{ThompsonMoranSwensonBook}). 
As an application, for example, the size of a gamma-ray burst (GRB) 
can be estimated using the interstellar scintillation of radio afterglow
of a GRB (See, e.g., \cite{Frail1997}).  On the other hand, 
there are compact gravitational wave sources whose linear sizes  
are of order the wavelength they emit. 
These include relativistic 
coalescing binaries such as neutron stars binaries and isolated
non-axisymmetric pulsars, both of which are two of the most promising 
sources (e.g., \cite{Cutler2002})  
for the gravitational wave detectors that are 
currently working or planned, such as 
LIGO \cite{LIGO1992}, 
LCGT \cite{LCGT1999}, 
LISA \cite{LISA1996}, 
BBO \cite{Phinney2003}, 
and 
DECIGO \cite{SetoKawamuraNakamura2001}.  
At around the BBO/DECIGO observing frequency of $f = 0.1$Hz,  
the orbital separation $l$ of a neutron stars binary would be 
$l \sim 10^4 \km (2 M_{\rm NS}/2.8 M_{\odot})^{1/3} (f/0.1\Hz)^{-2/3}$, 
while 
$l_0 \sim 4\times 10^{2} \AU (f/0.1 \Hz)^{-1} (\theta/10'')^{-1} >> $ $l$.

In the above argument, we saw that there is an upper limit on the size 
of the emitter to see interference. This is not a sufficient 
condition for interference pattern to form in the 
space and for us to observe its amplitude changing as the Earth (or a detector) 
moves in that spatial pattern. For a clear interference pattern to form in the 
space, coherence time must be larger than the time delay among multiple
images. Note that the coherence time is of order 
$1/\Delta f$ where $\Delta f$ is the frequency shift during
an observation. Suppose an extragalactic neutron stars binary as 
a gravitational wave source. Then 
to have a large enough 
signal to noise ratio with BBO/DECIGO to claim detection, we may
integrate the signal from the binary for a couple of years during 
which the signal frequency of the source binary would shift over of 
order of the BBO/DECIGO observing frequency, $f \sim \Delta f \sim 0.1$
Hz. The coherence time would then be $10$ seconds, which is smaller than, for example,    
the time delays of the known gravitational lensing systems 
listed in the CASTLE data base \cite{CASTLE,CASTLE1998}.   
So unless the time delay is smaller than $\sim 10$ seconds or the source
is highly monochromatic, we are not in a situation where 
we (on the Earth) move in a clear spatial interference pattern. 

The above argument does not preclude a possibility of detecting 
an interference term in the time domain, or equivalently, in the
frequency domain. In this paper, we do not aim to detect an interference
pattern in the space. Rather, we propose a method of a filtered 
cross-correlation of gravitational lens images in the frequency domain.
We shall point out that the interference term contains information 
of the relative transverse velocity of the source-lens-observer system
and study how well we could determine it using the future BBO/DECIGO
detector.  

There have already been several works that point out importance of 
interference and, more broadly, wave optics effects in  
gravitational lensing phenomena of gravitational 
waves \cite{Peterson1991,Takahashi2004,Matsunaga2006,Takahashi2006,Yoo2007}. 
Since a gravitational wave source such as a coalescing binary  
is essentially a point source, we have to use wave optics rather 
than geometrical optics approximation for such a source near 
caustics \cite{Schneider1992,Baraldo1999}.   
The works \cite{Takahashi2006,Yoo2007} 
studied observable effects on the gravitational waves propagating 
in an inhomogeneous universe. In this paper, we propose a method for 
extracting an interference term 
in the geometrical optics limit and for obtaining information on 
a relative transverse velocity of the source-lens-observer system. 
There are various methods to measure a relative transverse velocity 
of astronomical objects at a cosmological distance 
in the electromagnetic astronomy. 
Those includes for example, a method to measure transverse velocities of 
clusters of galaxies detecting a signature of gravitational scattering 
of the cosmic microwave background \cite{Birkinshaw1983,Pyne1993,Aso2002} and 
to measure those of galaxies using a parallax effect due to
gravitational micro-lensing \cite{Grieger1986,Gould1995}. 
Possibilities are also discussed in in the gravitational lens 
literature by \cite{Molnar2003}. 
Gravitational lensing of gravitational waves possibly offers another   
way to measure a relative transverse velocity of cosmological objects 
in addition to those methods above.

Our method could be applied to various sources including continuous waves 
from pulsars, cosmic strings and so on. However, to be specific, we
consider a coalescence neutron stars binary at a cosmological distance.

This paper is organized as follows. In the next section, we explain our
situation by discussing a (astronomical scale) Young's interference experiment.
In Sec. III, we remind 
readers of the basic equations and the notations 
for study of gravitational lensing of gravitational waves, for which 
we follow \cite{TakahashiNakamura2003}.  After remarks on
a situation we would be in when searching for gravitational lensing of 
gravitational waves and on the geometrical optics approximation, 
Sec IV explains how to extract interference term when the approximation 
applies. In the section \ref{sec:transversevelocity}, we 
propose a method for extracting information contained in the 
interference term, namely, the relative transverse velocity of  
the source-lens-observer system. We then study a correlation 
between the time delay and the transverse velocity in 
Sec \ref{sec:correlation_time_delay_and_transverse_velocity} 
and a correlation among parameters in 
\ref{sec:correlation_among_parameters} in the absence of noise. 
Sec. \ref{sec:accuracy} then shows the precision to which we 
could measure a transverse velocity by introducing detector noise. 
The section \ref{sec:summary} gives summary of our result. 

In this paper we use a unit of $G = c = 1$ unless otherwise written 
explicitly. An amplitude of a gravitational wave inversely depends on
the luminosity distance from the observer to the source, 
to compute which   
we assume a flat universe with 
$\Omega_{\Lambda} = 0.7$, $\Omega_{\matter} = 0.3$, and 
$h = 0.7$.

\section{Young's interference experiment}

To explain our situation, consider the (astronomical scale) 
Young's two pin-holes 
interference experiment. The (gravitational) wave 
intensity $I$ at the observer's time and
position $(t,x)$ on the screen is 
\begin{eqnarray}
I(t,x) &=& \left|\frac{1}{r_1}e^{i\sigma_1} + \frac{1}{r_2}e^{i\sigma_2}\right|^2
 \simeq 2\frac{1}{r^2}(1+\cos(\sigma_1-\sigma_2)),  
\end{eqnarray}
where $r_k$ is the distance from the $k$-th pin-hole to the
observer and $\sigma_k$ is the phase of the wave from the $k$-th pin-hole. 
We neglect the difference in $r_1$ and $r_2$ in the amplitude and write
$r_1 \simeq r_2 \equiv r$. 
The phase difference $\sigma(t,x) \equiv \sigma_1 - \sigma_2$ may be written as 
\begin{eqnarray}
\sigma(t,x) =  
2\pi \int_{t_r}^{t_r + \Delta t_d(t,x)} f(t')dt',
\end{eqnarray} 
where $t_r = t_r(t,x) = t - \tau_1(t,x)$ 
is the source retarded time for the wave through the pin-hole 1,
$\tau_k(t,x)$ is the time length for a wave crest to propagate from the source 
to the observer through the $k$-th pin-hole, and $\Delta t_d(t,x) = \tau_2(t,x)
- \tau_1(t,x)$ is the time delay 
between two pin-holes.

For a monochromatic wave at the frequency $f(t) =
f_c =$ constant, $\sigma(t,x) = 2\pi f_c \Delta t_d(t,x)$ and the scale of the 
interference pattern is $\Delta x = c/\theta/f/2$ ($\theta$ being the
angular separation of the two pin-holes seen from the observer) as
usual. Known possible monochromatic gravitational wave sources 
are isolated pulsars. The planned next generation
gravitational wave antennas would see those within the Milky way
galaxy. So considering micro-lensing phenomena on a gravitational wave   
emitted by 
an isolated millisecond pulsar in a globular cluster (say), the spatial
scale of the interference pattern is 
\begin{eqnarray}
\Delta x &\simeq&
3\times 10^{10} \km \left(\frac{1\mas}{\theta}\right)\left(\frac{1\kHz}{f_c}\right)
 \simeq 200 \AU. 
\label{eq:interferencePatternScale}
\end{eqnarray} 
Suppose, in the spatial interference pattern due to the monochromatic wave,  
we move at 200 km/s (about the orbital speed of the Solar system around
the Galactic center), it takes $\sim 5$ years for us to go across the 
spatial interference pattern. The wave intensity will vary in time as 
\begin{eqnarray}
I(t,x) \sim \cos\left(2\pi\frac{f v \theta}{c}t + \const. \right) + \const. 
\end{eqnarray} 

In principle, for a monochromatic wave, we would be able to detect 
relative transverse velocity of the source-lens-observer system by
using interference pattern (by measuring the temporal variation of $I(t,x)$). If we recognise such interference
pattern, we would conclude that gravitational force shows interference phenomena and
propagate as ``wave''. Other than that, the information we could
obtain, $v \theta$, is in principle the same as that in micro-lensing experiments in the
conventional electromagnetic astronomy. (And the number of millisecond 
pulsars is much smaller than that of the stars in the Magellanic clouds, say.)

When observing pulsars within the Milky way, it is also important to
note that diffraction effect becomes non-negligible when gravitational
wavelength $\lambda$ is larger than the lens mass $M_{L}$ 
\cite{TakahashiNakamura2003}, or $M_L \simlt 10^{2} M_{\odot}(f/1\kHz)^{-1}$. 
If we observe interference of gravitational waves from pulsars in the
Milky way, then the lens should have mass larger than $\sim 10^{2} M_{\odot}$

Eq. (\ref{eq:interferencePatternScale}) suggests a cosmological
application of gravitational lensing phenomena on gravitational waves.
\begin{eqnarray}
\frac{\Delta x}{v} &\simeq&
5 \years
\left(\frac{200\kms}{v}\right)\left(\frac{10''}{\theta}\right)\left(\frac{0.1\Hz}{f_c}\right).  
\end{eqnarray} 
Unfortunately for our purpose of seeing an interference pattern, 
the most promising cosmological sources, 
relativistic compact binaries in an inspiralling phase, do not emit 
monochromatic wave. For such a source the frequency
of the gravitational wave varies in time,  to the lowest order,   
\begin{eqnarray}
\frac{df}{dt} &=& \frac{96\pi^{8/3}}{5} \left(\frac{GM_{\rm c}}{c^3}\right)^{5/3}f^{11/3},
\end{eqnarray} 
with $M_{\rm c} = (m_1m_2)^{5/3}/(m_1+m_2)^{1/5}$ is the chirp mass,
$m_k$ being the mass of the $k$-th star in the binary.
For  wave having time-varying frequency $df(t)/dt = A f^{\eta}(t)$
with $A$ and $\eta$ constants, 
\begin{eqnarray}
\frac{\sigma(t,x)}{2\pi} &=& 
((\eta-1)(t_c-t-\Delta t_d(t,x))A + f_{\max}^{1-\eta} )^{\frac{1}{1-\eta}} 
-  
((\eta-1)(t_c-t)A + f_{\max}^{1-\eta} )^{\frac{1}{1-\eta}}, 
\end{eqnarray} 
with $f(t = t_c) = f_{\max}$ at which inspiralling phase ends and the
binary stars crush into each other. The interference pattern changes with time
even if there is no relative transverse velocity in the system. Moreover
because of the smallness of the amplitudes of possible gravitational
waves, we normally need to integrate signals for some time duration 
to get a sufficiently-large-for-claiming-detection signal to noise ratio. Schematically, we would do 
\begin{eqnarray}
\int^{T_{\obs}}_0 I(t',vt' + \const.) dt'.  
\end{eqnarray} 
$T_{\obs}$ is determined by either the source lifetime, its pass-time 
over the frequency band of our detector, or a pre-set threshold for claiming
detection (, say, signal to noise ratio larger than 5). 
Large $T_{\obs}$ in the integration makes the interference term vanishingly smaller than 
the bolometric (non-interference) term. For a massive compact binary, 
depending on binary masses, it
is possible that $T_{\obs} << (\time\,\,\delay)$ so that one wave packet
through the pin-hole 1 comes to the detector, and after some quiet period, the second 
through the pin-hole 2 visits it. In this case, no interference pattern
forms in the space and we would detect their interference by taking 
a cross-correlation (in a computer).
This paper shall use a filtered cross-correlation technique to take care of the issues above. 
But before moving onto explaining the technique we use in this paper, we 
start our discussion by  
explaining lensed waveform of gravitational waves in the next section.

\section{Lensed waveform of gravitational wave}

We consider a coalescence compact stars binary 
at the redshift $z_S$ 
with the redshifted chirp mass 
${ M}_{cz} = { M}_c(1+z_S)$ as a target source to be lensed. 
The detector is assumed to be a space borne interferometer such as 
(one-triangle) BBO/DECIGO, which outputs 
two independent data streams $h_{\alpha = I, II}(f)$.  
When the gravitational wave from the 
binary is lensed by a lensing object 
of redshifted mass $M_{Lz} = M_L(1+z_L)$, 
the wave through the $j$-th image to the detector  
in the frequency domain would be written, in the geometrical optics
limit and in the stationary phase approximation,  
as \cite{TakahashiNakamura2003}  
\begin{eqnarray}
h_{\alpha,j}^L(f) &=& \frac{\sqrt{3}}{2}|\mu_j|^{1/2}
\Lambda_{\alpha,j}(t_j) 
e^{-i\pi n_j -i\Phi_{\alpha,j}(t_j)}A f^{-7/6} e^{i\Psi_j(f)}, 
\label{eq:lensedwaveform}
\end{eqnarray} 
with
\begin{eqnarray}
\Phi_{\alpha,j}(t) &=& \phi_{D,j}(t) + \phi_{p,j,\alpha}(t), 
\end{eqnarray} 
where $n_j = 0, 1/2$ and 1 when the j-th image corresponds to a minimum, saddle, and
maximum point, respectively. 
$\Lambda_{\alpha,j}(t)$ is written 
in terms of  the detector beam pattern functions  
$F_{\alpha}^{+}(t),F_{\alpha}^{\times}(t)$ \cite{Cutler1998} and  
defined as 
\begin{eqnarray}
\Lambda_{\alpha,j}(t) &=& \left(
(1+ (\vec L\cdot \vec N_j)^2)^2 F_{\alpha}^+(t) 
+ 4 (\vec L\cdot \vec N_j)^2 F_{\alpha}^{\times}(t)  
\right)^{1/2},
\end{eqnarray} 
where $\vec L$ (given by $\bar \theta_L,\bar \phi_L$) and $\vec N_j$
(given by $\bar \theta_{S,j},\bar\phi_{S,j}$) are the direction unit vector 
of the binary orbital angular momentum and 
the direction unit vector toward the j-th image. 
These vectors are
defined in the fixed barycenter frame of the solar system.   
The detector phase $\phi_{D,j}(t)$  and the source phase
$\phi_{p,j,\alpha}(t)$ are 
\begin{eqnarray}
\phi_{D,j}(t) 
&=& 2\pi f(t) R \sin\bar \theta_{S,j}\cos\left(\bar \phi(t)
	- \bar \phi_{S,j}\right),  
\\
\phi_{p,j,\alpha}(t) &=& \tan^{-1}\left(\frac{2(\vec L\cdot\vec N_j)
				 F^{\times}_{\alpha}(t)}{(1+(\vec
				 L\cdot\vec N_j)^2)F^+_{\alpha}(t)}\right), 
\end{eqnarray} 
where $\bar \phi(t) = 2\pi t/T + \bar \phi_0$ with $R = 1\AU$ and $T=$ 1 year for 
BBO/DECIGO (Those detectors are planned to orbit around the solar system
barycenter with its average orbital radius 1 AU. 
For simplicity we assumed a circular orbit for the detector here).
The amplitude of the gravitational wave $A$  and the phase of the 
gravitational wave $\Psi_j(f)$, to the Newtonian approximation, are 
\begin{eqnarray}
\Psi_j(f) &=& 2\pi f t_{d,j} + \Psi(f)= 
2\pi f t_{c,j} - \phi_c - \frac{\pi}{4} 
+ \frac{3}{4}\left(8\pi{ M}_{cz}f\right)^{-5/3},
\label{eq:PSIofF}
\\ 
A &=& \frac{1}{D_S(1+z_S)^2}
\left(\frac{5}{96}\right)^{1/2}\pi^{-2/3}{ M}_{cz}^{5/6},
\end{eqnarray} 
where $D_S$ is the source angular diameter distance. 
$\mu_j$ and $t_{d,j}$ 
are 
the amplification factor and 
the time delay for the j-th image measured with respect to the fictitious 
time of arrival of non-lensed signal. 
$t_{c,j} = t_c + t_{d,j}$.  
$t_c$ and $\phi_c$ are the time and the phase of the binary 
coalescence.
Finally, the time variable $t$ in the above equations 
must be understood to be a function 
of frequency as 
\begin{eqnarray}
t(f) &=& = t_{c} - 5(8\pi f)^{-8/3}{ M}_{cz}^{-5/3},
\\
t_j(f) &=& t(f) + t_{d,j} = t_{c,j} - 5(8\pi f)^{-8/3}{ M}_{cz}^{-5/3}.
\label{eq:TofF}
\end{eqnarray} 

In a gravitational wave search, the unknown parameters would be,  
to the lowest order, 
\begin{eqnarray}
\{ 
{ M}_{cz},  
\phi_{c},  
t_{c},
D_S,
\bar \theta_{S,j},      
\bar \phi_{S,j},
\vec L,
\kappa_1 - \kappa_2,   
\Delta t_d = t_{d,2} - t_{d,1},
|\mu_1|/|\mu_2|
\},
\label{eq:list_of_search_parameters}
\end{eqnarray}  
where the last three depend on the lens property.

\section{Search for a lensed gravitational wave}
\label{sec:explaning_situation}

We consider the following situation. First of all, for simplicity 
we study that two images occur due to gravitational lensing. 
Then the wave from one lens image would
be in the detector's observation frequency band in the time duration
from, say, $t_{s,1}$ to $t_{e,1}$.  The
second wave would be in the band after the time delay $\Delta t_d$ later from 
$t_{s,2} = t_{s,1} + \Delta t_d$ to $t_{e,2} = t_{e,1} + \Delta t_d$.  We assume the 
observation frequency and the time delay satisfy $f
\Delta t_d \sim 0.1 (f/0.1\Hz) \Delta t_d >> 1$ so that we can use the geometrical optics 
approximation (\cite{TakahashiNakamura2003} studied the case where wave optics is important). 
As as result, we may detect the two waves separately using 
unlensed waveform templates as if those were unlensed, in the same sky direction  
(The angular resolution of the gravitational wave source is $\sim
10^{-2}$(Signal to noise ratio)$^{-1}$  radians 
with the current BBO design \cite{CutlerHarms2003}, so we do not expect to resolve images).  
We would then realize that those two waves have similar parameters sets  
${ M}_{cz},\phi_c,\vec L$, and conclude that those two are actually 
due to one lensed event. We may find the (relative) time delay 
$\Delta t_{d} = t_{d,2} - t_{d,1}$ and the ratio of the 
 magnification $|\mu_1|^{1/2}/|\mu_2|^{1/2} = A_1/A_2$.                        
The observational errors of these two numbers are at the same level as those 
of $t_c$ and $A$. We might now cross-correlate the two waves to see 
interference between these two (in a computer). However, a simple 
cross-correlation disappears when the geometrical optics approximation
applies (and this is the reason we could detect two waves separately, anyway).  
Let us see this using some equations below and propose a  method 
to extract information contained in the interference term even 
in this case.

\subsection{How to extract the interference term when the 
geometrical optics approximation applies}
\label{sec:howto}

We consider a sort of cross-correlation between the two waves, given by 
(\ref{eq:lensedwaveform}). We use the waves with the supports in the
time domain assumed to be from $t_{e,1} - \Delta t_d$ to $t_{e,1}$  for the wave that 
reaches the detector first, and from $t_{e,2} - \Delta t_d = t_{e,1}$ to
$t_{e,2}$ for the second. The detector's output $s_{k}(t)$ in the two time
segments ($k=A,B$) are 
\begin{eqnarray}
s_A(t) &=& \sum_j h_j(t) \,\,\, {\rm for} \,\,\, t_{e,1}-\Delta t_d \le t \le t_{e,1}, \\
s_B(t) &=& h_2(t) \,\,\, {\rm for} \,\,\, t_{e,1} \le t \le t_{e,2}. 
\label{eq:detecter_output_without_noise}
\end{eqnarray} 
The Fourier expansions of the gravitational wave $h_1(t)$ in the time
segment $A$ and that of $h_2(t)$ in the time segment $B$ 
have a support from $f_i$ to $f_e$ in the frequency domain. The
frequency domain support of $h_2(t)$ in the time segment $A$ is lower than 
$f_i$ . We
now compute a simple cross-correlation between the two waves as 
\begin{eqnarray}
4 Re \int_{f_i}^{f_e}
\frac{s_{A}(f) s_{B}^*(f)}{S_h(f)}df
&=& 
4 Re \sum_{\alpha} 
\int_{f_i}^{f_e}
\frac{h_{\alpha,1}^L(f) h_{\alpha,2}^{L*}(f)}{S_h(f)}df,
\label{eq:detecter_output_cross_correlation_without_noise}
\end{eqnarray} 
with $S_h$ the one-sided spectral density of the detector's noise. 
Because we defined the integral region from $f_i$ to $f_e$ so that
there is no auto-correlation term of $h_{\alpha,2}^L(f)$ which otherwise
appears due to $h_2(t)$ term in $s_A(t)$. Now,
$h_{\alpha,j}^L(f)$ is given by Eq. (\ref{eq:lensedwaveform}) and the
simple cross-correlation becomes 
\begin{eqnarray}
\lefteqn{
4 Re \int_{f_i}^{f_e}
\frac{h_{\alpha,1}^L(f) h_{\alpha,2}^{L*}(f)}{S_h(f)}df}
\nonumber \\
&=& 
\frac{3}{2}|\mu_1|^{1/2}|\mu_2|^{1/2}
A^2 
\nonumber \\
&\times&
Re
\int_{f_i}^{f_e}\frac{f^{-7/3}}{S_h(f)}
\Lambda_{\alpha,1}(t_1)\Lambda_{\alpha}(t_2)
e^{2\pi i f \Delta t_d - i \pi (\kappa_1 - \kappa_2)}
e^{i\Phi_{\alpha,2}(t_2) -i\Phi_{\alpha,1}(t_1)}
df.  
\end{eqnarray} 
This integrand is oscillatory and the result is effectively zero 
if $|f \Delta t_{d}| >> 1$.

Now, how can we extract non-zero cross-correlation? We propose a
filtered cross-correlation statistic. We multiply the two images, and before the frequency domain 
integration we further
multiply $\cos( 2\pi f \Delta t^T_{d} + \Theta(f,{\mathbf p^T}))$  
where $\Delta t_d = t_{d,2}-t_{d,1}$,  
the tuplet 
$\mathbf p = (t_{c,1},t_{c,2},M_{cz},\kappa_1-\kappa_2,\vec N,\vec L)$,  
\begin{eqnarray}
\Theta(f,\mathbf p) 
\equiv  
\Phi_{\alpha,2}(t_2(f)) - \Phi_{\alpha,1}(t_1(f)) 
+ \pi(\kappa_1-\kappa_2),  
\label{eq:define_Theta}
\end{eqnarray} 
and $\mbox{}^T$ denoting a template.
We define our detection statistic $\zeta_{\alpha}$ as 
\begin{eqnarray}
\zeta_{\alpha}
&=& 4 Re \int_{f_i}^{f_e}
\frac{h^L_{\alpha,1}(f)h^{L*}_{\alpha,2}(f)}{S_h(f)}
\cos(  2\pi f \Delta t^T_{d} + \Theta(f,{\bf p}^T)) 
df
\nonumber \\
&\simeq& 
3|\mu_1|^{1/2}|\mu_2|^{1/2}A^2
\nonumber \\
&\times&
\int_{f_i}^{f_e}\frac{f^{-7/3}}{S_h(f)}
\Lambda_{\alpha,1}(t_1)\Lambda_{\alpha,2}(t_2)
\cos(  2\pi f (\Delta t_{d}-\Delta t_{d}^T) + 
\Theta(f,\mathbf p) - \Theta(f,{\mathbf p^T}))
df.  
\label{eq:zeta_statistic}
\end{eqnarray} 
Maximizing an absolute value of the detection statistic $|\zeta|$ (where 
$\zeta \equiv \sum_{\alpha}\zeta_{\alpha}$), we obtain our estimate of 
the time delay $\Delta t_d$.  
In the next section 
we study an effect on our statistic $\zeta$ of a relative transverse velocity of the source-lens-observer
system.  In the later sections, we shall study precision to which we could
measure $\Delta t_d$ and the transverse velocity and correlations among
errors of the parameters $\Delta t_d$, the transverse velocity, and ${\bf p}$.

\section{Transverse velocity}
\label{sec:transversevelocity}

We can measure 
$\{t_{c,j},{ M}_{cz}, \bar \theta_S,\bar \phi_S, \vec L, n_j \}$ 
using gravitational lensing phenomena in the usual electromagnetic 
astronomy.  
When one can use information in an interference term, we could get 
new information on the lens/source object: its relative transverse 
velocity. When there is a relative velocity among the
source-lens-observer, there may be two effects. One is the time
variation of the direction of the images $\vec N_j$, which may not 
be detected for cosmological sources 
in the foreseeable future because of the poor pointing 
ability of the planned gravitational wave detectors. The other 
is the Doppler effect which causes rescaling of the mass of the 
source, the time variables and the amplitudes of the 
gravitational waves.  
The rescaling due to the Doppler effect 
differs for different images as  
\begin{eqnarray}
\tilde t_{jk}(f) &=& \tilde t_{c,jk} 
- 5(8\pi f)^{-8/3} \tilde M_{cz,k}^{-5/3},  
\end{eqnarray} 
and 
$
\tilde{ M}_{cz,j} = \gamma_j { M}_{cz}, 
\tilde t_{c,jk} = \gamma_k t_{c,j},
$
and 
$ 
\tilde A_j = \gamma_j^{- 7/6}  A 
$
with
$
\gamma_j = (1 + \vec N_j\cdot \vec \beta), 
$
where the velocity $\vec \beta = \vec v /c$ is a linear combination of  
the observer velocity $\vec v_O$,  the lens velocity $\vec v_L$, 
and the source velocity $\vec v_S$ as \cite{Kayser1986,Wicknitz2004}
\begin{eqnarray}
\vec v &=& 
\vec v_O + \frac{1+z_L}{1+z_S}\frac{D_L}{D_{LS}}\vec v_S - \frac{D_{LS} + D_L}{D_{LS}}\vec
v_L,  
\label{eq:transverse_velocity}
\end{eqnarray} 
with $D_{LS}$ being the angular diameter distance between the lens and
the source.

The lensed gravitational waveform from a coalescing binary now becomes   
\begin{eqnarray}
\tilde h_{\alpha,j}^L(f) &=& 
\frac{\sqrt{3}}{2}
|\mu_j|^{1/2}\Lambda_{\alpha,j}(\tilde t_{jj})
e^{- i \pi n_j}
e^{-i\Phi_{\alpha,j}(\tilde t_{jj})}
\tilde A_j f^{-7/6} e^{i\tilde \Psi_{jj}(f)}, 
\end{eqnarray} 
with 
\begin{eqnarray}
\tilde \Psi_{jk}(f) &=& 
2\pi f \tilde t_{c,jk} - \phi_c - \frac{\pi}{4} 
+ \frac{3}{4}\left(8\pi\tilde { M}_{cz,k}f\right)^{-5/3}.  
\end{eqnarray} 
If there is only one image, then we could not resolve the degeneracy and 
would get, in principle, a biased estimate of the masses and the
source distance. However, if we know 
in advance 
that we see multi-waves due to gravitational lensing 
(see Sec. \ref{sec:explaning_situation}), 
then we could find the differences of the Doppler factors among different images.  

For later convenience, we 
introduce a notation for the ratio of the Doppler factors as 
\begin{eqnarray}
\Gamma &=& 
\frac{\gamma_1}{\gamma_2} 
 = 
1 + (\vec N_1 - \vec N_2)\cdot\vec\beta + O(\beta^2).
\end{eqnarray} 
Note that $\Gamma - 1 \simeq 
\theta \beta_{\perp}  
$ 
where $\theta$ is the angular size of the lens. 
Finding $\Gamma$ using the interference term of 
the lensed gravitational wave then gives us 
the transverse velocity $\beta_{\perp}$
\footnote{More precisely, 
$\beta_{\perp}$ is the relative velocity of the source-lens-observer
system perpendicular 
to the line of the sight and along the 
connection vector between the two images.}, given that $\theta$ is known
from, say, electromagnetic observation of the lens object
(say, galaxies/clusters) and the host
object (say, a galaxy) of the gravitational wave source.

We compute a cross-correlation between the lensed wave with itself but with frequency shifted by $\Gamma^T$ here 
$\mbox{}^T$ denoting a template, and further multiply the cosine filter as
in the previous section,   
\begin{eqnarray}
\zeta_{\alpha} &=& 
4 Re \int_{f_i}^{f_e}
\frac{df}{S_h(f)}
\tilde h^L_{\alpha,1}(f)
\tilde h^{L*}_{\alpha,2} \left(\Gamma^T f \right)
\cos(  2\pi f \Delta \tilde t^T_{d,1} + \Theta(f,\mathbf p^T))) 
\nonumber \\
&\simeq&
3 
|\mu_1|^{1/2} |\mu_2|^{1/2}\tilde A_1 \tilde A_2
Re \int_{f_i}^{f_e}
\frac{df}{S_h(f)}
f^{-7/3} 
{\cal H} 
\cos(  2\pi f \Delta \tilde t^T_{d,1} 
+ \Theta(f,\mathbf p^T))). 
\label{eq:transverseSNR}
\end{eqnarray} 

The function 
${\cal H}$ is 
defined 
as 
\begin{eqnarray}
{\cal H}
&=& 
\Lambda_{\alpha,1}(\tilde t_{11}(f))
\Lambda_{\alpha,2}(\tilde t_{22}(\Gamma^T f))
\nonumber \\
&\times&
e^{i\Phi_{\alpha,2}(\tilde t_{22}(\Gamma^T f)) 
 - i\Phi_{\alpha,1}(\tilde t_{11}(f))}
e^{i\tilde \Psi_{11}(f) - i \tilde \Psi_{22}(\Gamma^T f)}, 
\end{eqnarray}
with 
\begin{eqnarray}
\tilde \Psi_{11}(f) - 
\tilde \Psi_{22}(\Gamma^T f) &=& 
2\pi f \tilde t_{c,11}  
+ \frac{3}{4}\left(8\pi \tilde { M}_{cz,1}f\right)^{-5/3}
\nonumber \\
&&- 
2\pi f \Gamma^T \tilde t_{c,22}  
- \frac{3}{4}\left(8\pi \Gamma^T \tilde { M}_{cz,2}f\right)^{-5/3}.
\end{eqnarray} 
The parameters to be searched for are, in addition to the parameters $\mathbf p^T$, 
the time delay between images, $\Delta t_{d}$, and 
the $\Gamma$ parameter. When $\Gamma^T 
\simeq 1 + \theta\beta_{\perp} + \delta(\theta\beta_{\perp})$ 
with $\theta\beta_{\perp} << 1$ 
and $\delta (\theta\beta_{\perp}) << 1$, 
\begin{eqnarray}
\tilde \Psi_{11}(f) - 
\tilde \Psi_{22}(\Gamma^T f) 
&\simeq& 
2\pi f \Delta \tilde t_{d,1}  
- 
2\pi f \tilde t_{22}(f) \delta(\theta\beta_{\perp}),  
\label{eq:velocity_phase1}
\end{eqnarray} 
with $\Delta \tilde t_{d,1} = \gamma_1 \Delta t_{d}$.

Eq. (\ref{eq:velocity_phase1}) tells us that larger the observation
frequency $f$ is, and/or longer the observation time is, 
with a better precision we measure the time delay $\Delta t_d$ and/or 
$\Gamma$ \footnote{In fact, we measure 
$\Delta \tilde t_{d,1}$ (or $\Delta \tilde t_{d,2}$ depending on 
the filter one chooses) 
because we can detemine $\Gamma = \gamma_1/\gamma_2$ but 
cannot determine $\gamma_1$ (or $\gamma_2$) separately. 
For notational simplicity and from the fact that 
$\Delta \tilde t_{d,1} \simeq \Delta t_{d}$, we 
say we measure $\Delta t_{d}$ in the following sections.}. 
This latter tendency will be studied further in the
sections \ref{sec:correlation_among_parameters} and 
 \ref{sec:accuracy}. Eq. (\ref{eq:velocity_phase1}) also indicates a
 correlation between  
$\Delta t_d$ and $\Gamma$, which we will study in the next section.

\section{Correlation between the time delay and the transverse  velocity}
\label{sec:correlation_time_delay_and_transverse_velocity}

When one uses usual matched filtering technique to unlensed
gravitational waves, the search parameters are 8 in the lowest order (See the list
(\ref{eq:list_of_search_parameters})). To find the transverse velocity
of the source, in addition to those,  we search for  
two parameters: the time delay $\Delta t_{d}$ and the transverse velocity
$\Gamma$. Since these two parameters are the new aspects in our paper, in this section we show how 
our detection statistic 
$\zeta$ (where $\zeta \equiv \sum_{\alpha}\zeta_{\alpha}$) depends on 
these two  parameters.

\begin{center}
\begin{figure}
\caption{
$\zeta$ contour map in the $\Gamma^T - \Gamma$ and $\Delta t_d^T -
 \Delta t_d$ plane, in the absence of noise. 
The source, lens, and detector parameters are listed in the table
 \ref{tbl:Q0957+561}. All the template parameters other than $\Gamma^T$ and 
$\Delta t_d^T$ are fixed to be the right values.
The dotted line is
 Eq. (\ref{eq:correlation_td_Gamma}), 
which approximately follows the ridge of $\zeta = \zeta(\Gamma^T,\Delta t_d^T)$.
}
\centering\includegraphics[width=16cm]{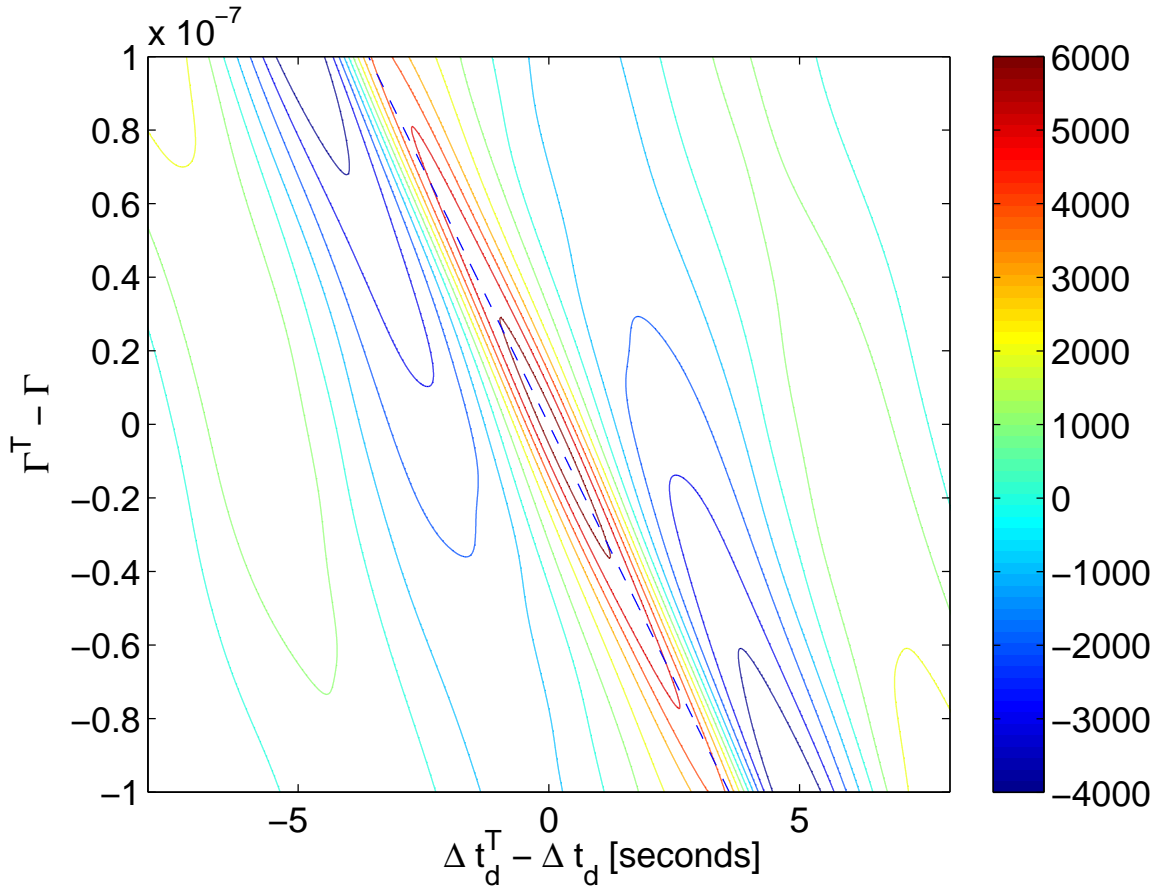}
\label{fig:Q_contour}
\end{figure} 
\end{center}

\begin{center}
\begin{figure}
\caption{
$\zeta$ along the line Eq. (\ref{eq:correlation_td_Gamma}) projected
 onto the $\Gamma^T - \Gamma$ axis, 
in the absence of noise. 
The source, lens, and detector parameters are listed in the table
 \ref{tbl:Q0957+561}. All the template parameters other than $\Gamma^T$ and 
$\Delta t_d^T$ are fixed to be the right values.
}
\centering\includegraphics[width=16cm]{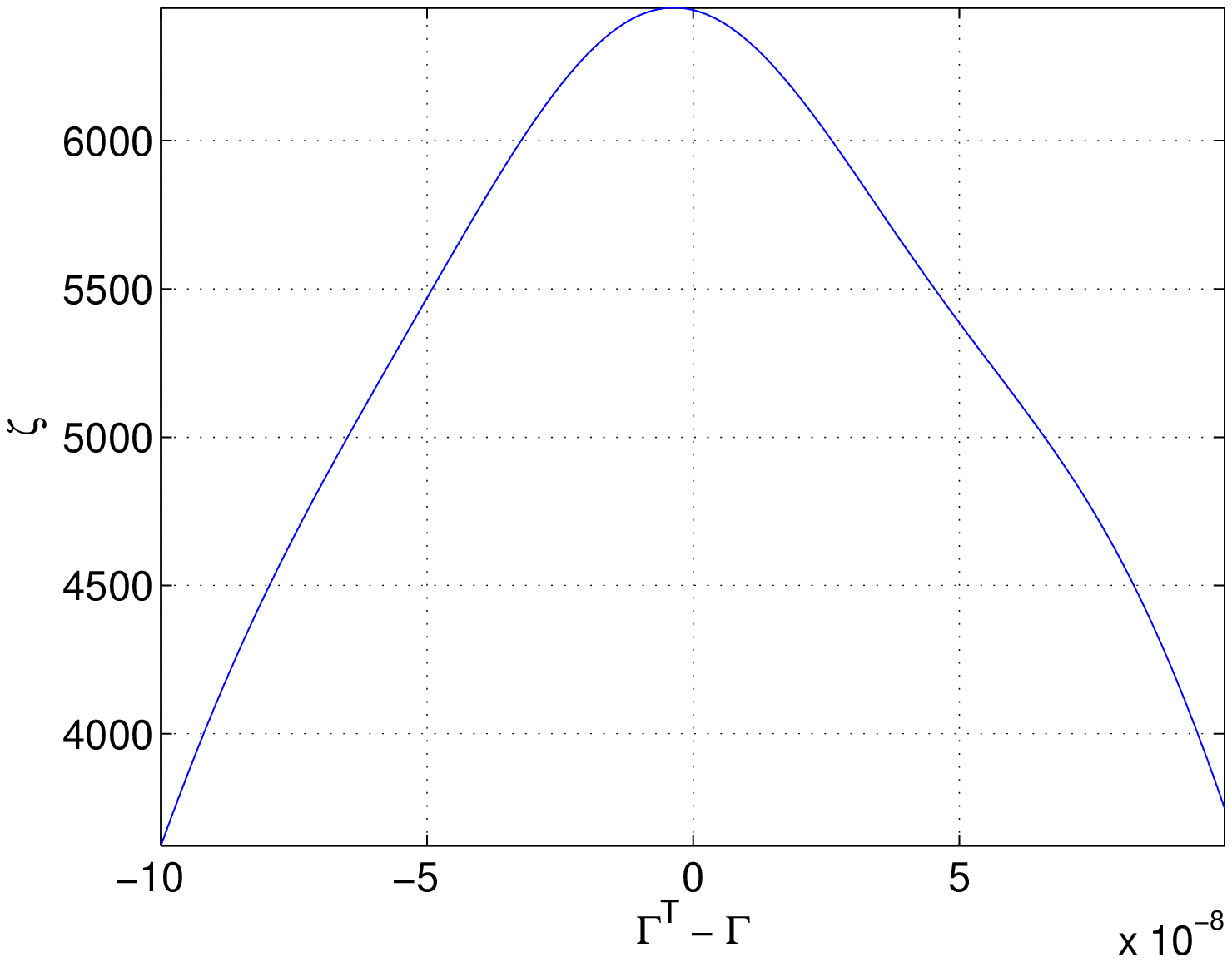}
\label{fig:Q_section}
\end{figure} 
\end{center}

To demonstrate a correlation between $\Delta t_d$ and $\Gamma$, 
we consider a source with
the parameters listed in the table \ref{tbl:Q0957+561}. 
The source redshift $z_S$, the lens redshift $z_L$, the image 
separation $\theta$, the
time-delay and the direction of the source are taken from 
the  
famous lens system Q0957+561 \cite{Walsh1979Natur,CASTLE}. With those
parameters, we compute our statistic $\zeta$ in the $\Gamma^T$-$\Delta t_d^T$
plane with other parameters fixed to be the right values.
Fig. \ref{fig:Q_contour} shows the contour map of $\zeta$, which clearly
shows a strong correlation between $\Gamma^T$ and $\Delta t_d^T$.
As can be inferred from Eq. (\ref{eq:velocity_phase1}), the correlation is found well approximated by 
\begin{eqnarray}
\Delta t_d^T -  \Delta t_d &=& - (\Gamma^T - \Gamma) T_{\obs},  
\label{eq:correlation_td_Gamma}
\end{eqnarray} 
near $\Delta t_d^T = \Delta t_d$. In fact, along this line and projected
onto $\Gamma$, the detection statistic behaves as in
Fig. \ref{fig:Q_section}, which shows the maximum appears at $\Gamma^T
\simeq \Gamma$: The maximum of $\zeta$ occurs slightly shifted
$\Gamma^T$ value from the true value ($\sim 1\%$).  This
happens because in Eq. (\ref{eq:transverseSNR}), we should have
used instead of $\Theta(f,{\bf p})$ defined in
Eq. (\ref{eq:define_Theta}), $\tilde \Theta(f,{\bf p})$ of the form 
\begin{eqnarray}
\tilde \Theta(f,\mathbf p) = 
\Phi_{\alpha,2}(t_2(\Gamma f)) - \Phi_{\alpha,1}(t_1(f)) 
+ \pi(\kappa_1-\kappa_2).
\end{eqnarray} 
Unfortunately, since use of this $\tilde \Theta$ demanded large computational
power in our code, we used $\Theta$ instead paying $\sim 1\%$ bias in
$\Gamma - 1$.

\begin{center}
\begin{table}
\caption{The source, lens, detector parameters for the figures 
\ref{fig:Q_contour} and \ref{fig:Q_section}. 
Here $T_{\obs}$ is the integration time (time for the wave to pass the
 frequency band from $f_i$ Hz 
 to $f_e$ Hz) in years, and  
$v = |\vec v|$: magnitude of the relative transverse velocity of the source-lens-observer
 system defined in Eq. (\ref{eq:transverse_velocity}). Other parameters are explained in the main body of the text. 
The source redshift $z_S$, the lens redshift $z_L$, the image separation $\theta$, the
time-delay and the direction of the source are taken from the  
lens system Q0957+561 \cite{Walsh1979Natur,CASTLE}. 
}
\begin{tabular}{ccccccc}
\hline
$f_i$ [Hz] & 
$f_e$ [Hz] & 
$T_{\obs}$ [yrs] &
$\Delta t_d$ [yrs] & 
$v$ [km/s] & 
$\theta$ [$''$] &
$\Gamma - 1$ 
\\ \hline 
$0.113$  & 
$1$  & 
$1.137$  &
$1.14$  & 
$480$  & 
$6.26$ &
$4.9\times10^{-8}$ 
 \\ \hline  \hline 
$\mu_1,\mu_2$ &
$t_c$ [yrs] & 
$m_1,m_2$ [$M_{\odot}$] &
$\bar \theta_L$ [rad] & 
$\bar \phi_L$ [rad] & 
$\bar \theta_S$ [rad] &
$\bar \phi_S$ [rad] 
\\ \hline  
$1.5,1$ &
$1.14$ &
$1.4,1.4$ &
$1.09$ &
$2.90$ &
$2.62$ &
$0.99$ 
\\ \hline  \hline 
$\bar \phi_0$ [rad] & 
$\phi_c$ [rad] &  
$\alpha_0$ [rad] &
$z_S$ &
$z_L$ 
\\ \hline  
$2.64$ & 
$2.55$ &  
	 $2.06$ &
$1.41$ & 
$0.36$ & 
 \\ \hline  
\label{tbl:Q0957+561}
\end{tabular}
\end{table} 
\end{center}

\section{Correlation among the parameters}
\label{sec:correlation_among_parameters}

As is mentioned in Sec. \ref{sec:explaning_situation}, we specifically
assume the following situation: Two images occur due to
gravitational lensing. A gravitational wave is in the
detector's observation frequency band during the time segment from 
$t_{s,1}$ to $t_{e,1}$. We use matched filtering technique to estimate 
the source parameters (8 in the lowest order,
Eq. (\ref{eq:list_of_search_parameters})). Another wave reaches the
detector from $t_{s,2} = t_{s,1} + \Delta t_d$ to $t_{e,2} = t_{e,1} + \Delta t_d$, and we 
use matched filtering technique to estimate the source parameters. From 
 similarity of the estimated values of the parameters, we would conclude 
that we see gravitational lensing of gravitational waves. We then
construct a filter $\Theta(f,{\bf p}^T)$ from the estimated parameters, 
compute a filtered cross-correlation $\zeta$,  and estimate $\Gamma$ and 
$\Delta t_d$. Therefore, in practice, an estimates of $\Gamma$ and
$\Delta t_d$ should
be affected by errors in estimates of other parameters ${\bf p}^T$ in 
$\Theta(f,{\bf p}^T)$ (We do not estimate ${\bf p}^T$ at the same time
as $\Gamma^T$ and $\Delta t_d$ when computing the filtered cross-correlation). 

\begin{center}
\begin{figure}
\caption{
A scatter plot of maximum locations of $\zeta$ in the $\Gamma^T - \Gamma$ and $\Delta t_d^T -
\Delta t_d$ plane for 1100 sets of randomly chosen source,
 lens, and detector parameters. Here $\Delta t_d = 1.14$ years. 
Symbols indicate $\zeta$ values: 
$\zeta_{\max} \sim 2.8\times 10^4$ denoting the maximum value of $\zeta$ in the 1100 trials,
diamonds for $0 \le \zeta < 0.25 \zeta_{\max}$, 
crosses for $0.25\zeta_{\max} \le \zeta < 0.5 \zeta_{\max}$, 
circles for $0.5\zeta_{\max} \le \zeta < 0.75 \zeta_{\max}$, and 
squares for $0.75\zeta_{\max} \le \zeta \le \zeta_{\max}$.
See Sec. \ref{sec:correlation_among_parameters} for details. 
}
\centering\includegraphics[width=16cm]{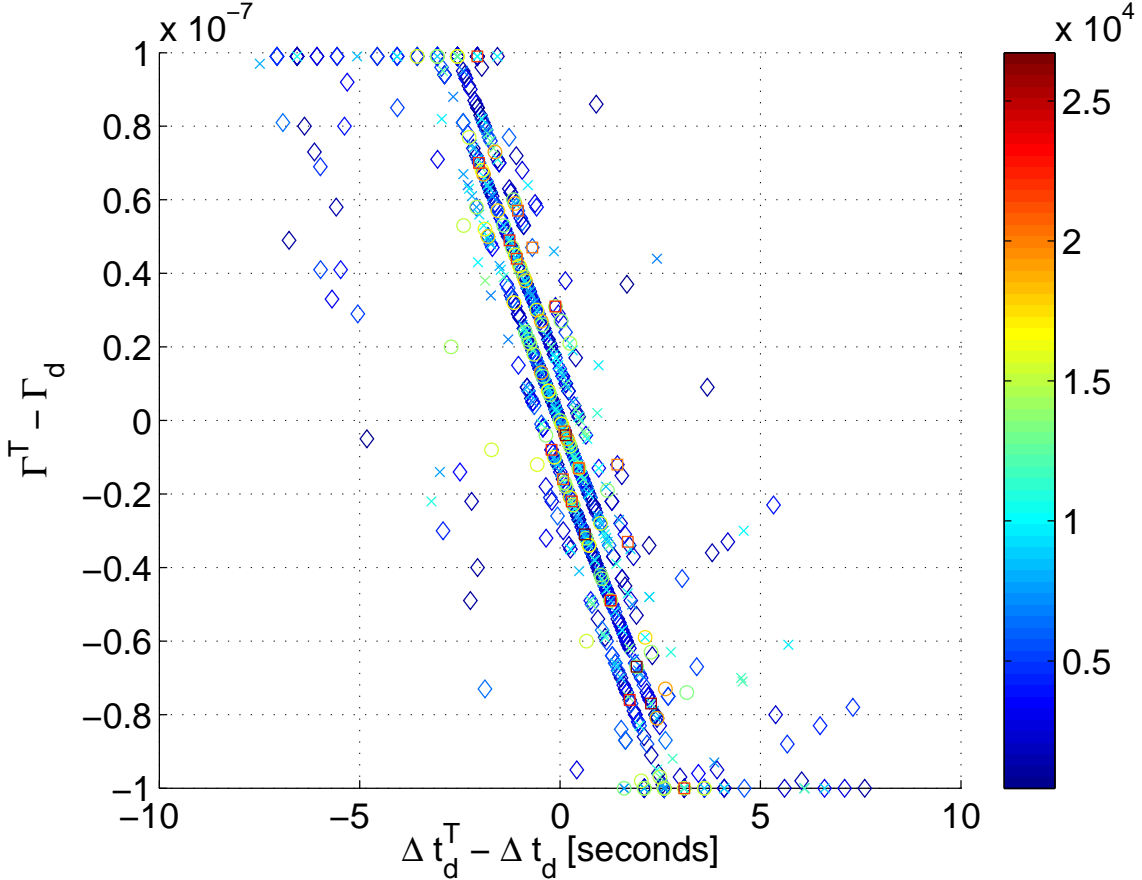}
\label{fig:td_Gamma_scatter_fmin0113}
\end{figure} 
\end{center}

\begin{center}
\begin{figure}
\caption{
The same as Fig. \ref{fig:td_Gamma_scatter_fmin0113} but 
here $\Delta t_d = 5.1$ years and correspondingly $f_i = 0.065$ Hz.
See Sec. \ref{sec:correlation_among_parameters} for details. 
}
\centering\includegraphics[width=16cm]{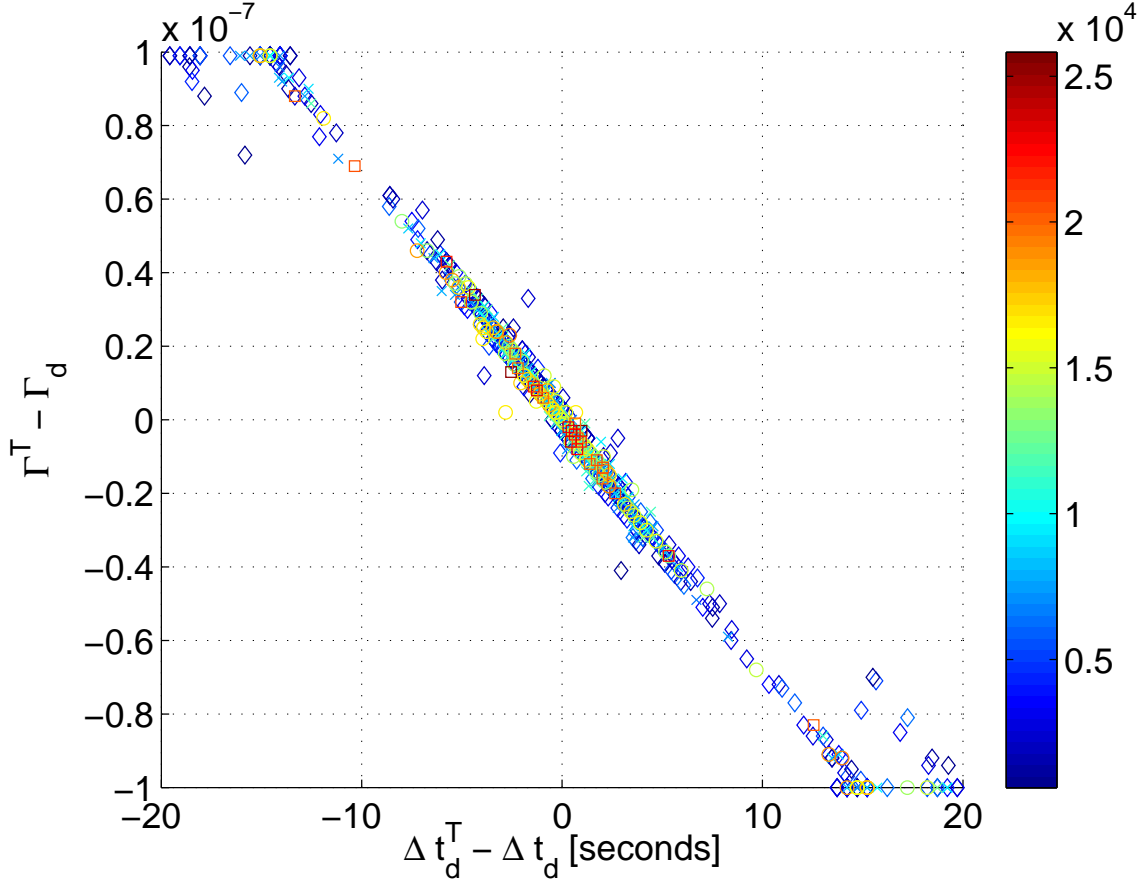}
\label{fig:td_Gamma_scatter_fmin0065}
\end{figure} 
\end{center}

To find the effect of the errors in ${\bf p}^T$
to the estimate of $\Gamma$, we performed the following simulations.  
We generate a wave with the source parameters 
${\bar \theta_L,\bar \phi_L,\bar \theta_S,\bar \phi_S,\phi_c,\bar \phi_0,\alpha_0}$ randomly
chosen (and thus we generate waves with randomly chosen ${\bf p}$). 
Other source parameters are fixed and listed in the table
\ref{tbl:Q0957+561}. For each parameters set, 
we take the template parameters by adding random Gaussian 
errors $\delta {\bf p}$ to the injected signal parameters ${\bf p}$ as 
${\bf p^T} = {\bf p} + \delta {\bf p}$.
The standard deviations of the errors are $10^{-6}$ for the relative errors
in the mass $m_j$ and the time of coalescence $t_c$, and $0.1$ radians 
for the angular parameters 
${\bar \theta_L,\bar \phi_L,\bar \theta_S,\bar \phi_S,\phi_c,\bar \phi_0,\alpha_0}$ \cite{Crowder2005}. 
We then
compute our statistic $\zeta$ and find $\Gamma$ and $\Delta t_d$ that
maximize $|\zeta|$. We repeated the above steps 1100 times. 
Fig. \ref{fig:td_Gamma_scatter_fmin0113} shows 
the result of the simulation. Here we search over the region of 
$|\Gamma^T - \Gamma| < 10^{-7} \cap 
|\Delta t_d^T - \Delta t_d| < 20$ sec (The size of the search region is limited from our
computer power). We see the estimates of $\Gamma$ and $\Delta t_d$
roughly satisfy Eq. (\ref{eq:correlation_td_Gamma}). Several points ($8 
\%$ of all) are
accumulated around the upper left and the lower right boundaries of the
search region, for which the estimates
should in fact be outside of the region.
Fig. \ref{fig:td_Gamma_scatter_fmin0065} shows the result 
of the same simulation but with the true time delay $\Delta t_d$ enlarged to $5.1$ years, and
correspondingly $f_i = 0.065$ Hz. We see a better correlation to
Eq. (\ref{eq:correlation_td_Gamma}) and a smaller points accumulated
around the boundary of the region ($3\%$). Both of the figures show a
tendency that
larger the value of $\zeta$ is, better the accuracy of our estimates for
$\Gamma$ and $\Delta t_d$ are.

\begin{center}
\begin{figure}
\caption{
Cumulative probability distributions of $|\Gamma^T - \Gamma|$. 
See Sec. \ref{sec:correlation_among_parameters} for details. 
}
\centering\includegraphics[width=16cm]{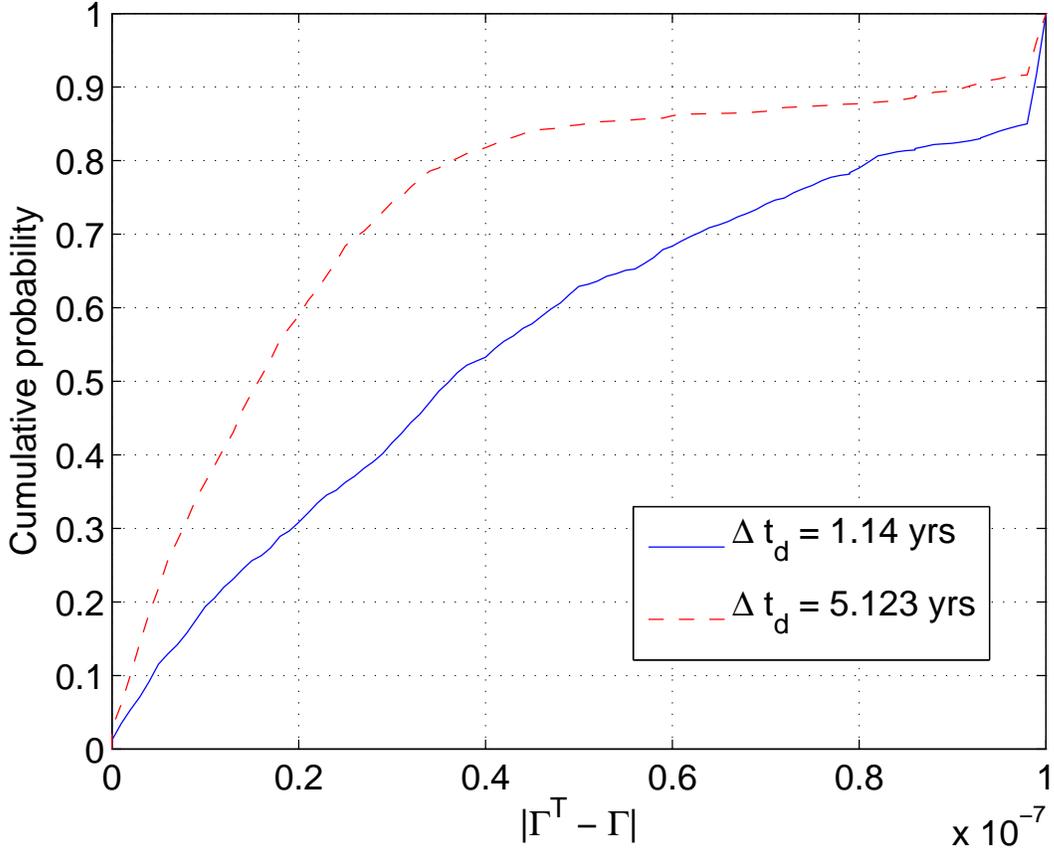}
\label{fig:ecdf_Gamma}
\end{figure} 
\end{center}

\begin{center}
\begin{figure}
\caption{
Cumulative probability distributions of $|\Delta t_d^T - \Delta t_d|$. 
See Sec. \ref{sec:correlation_among_parameters} for details. 
}
\centering\includegraphics[width=16cm]{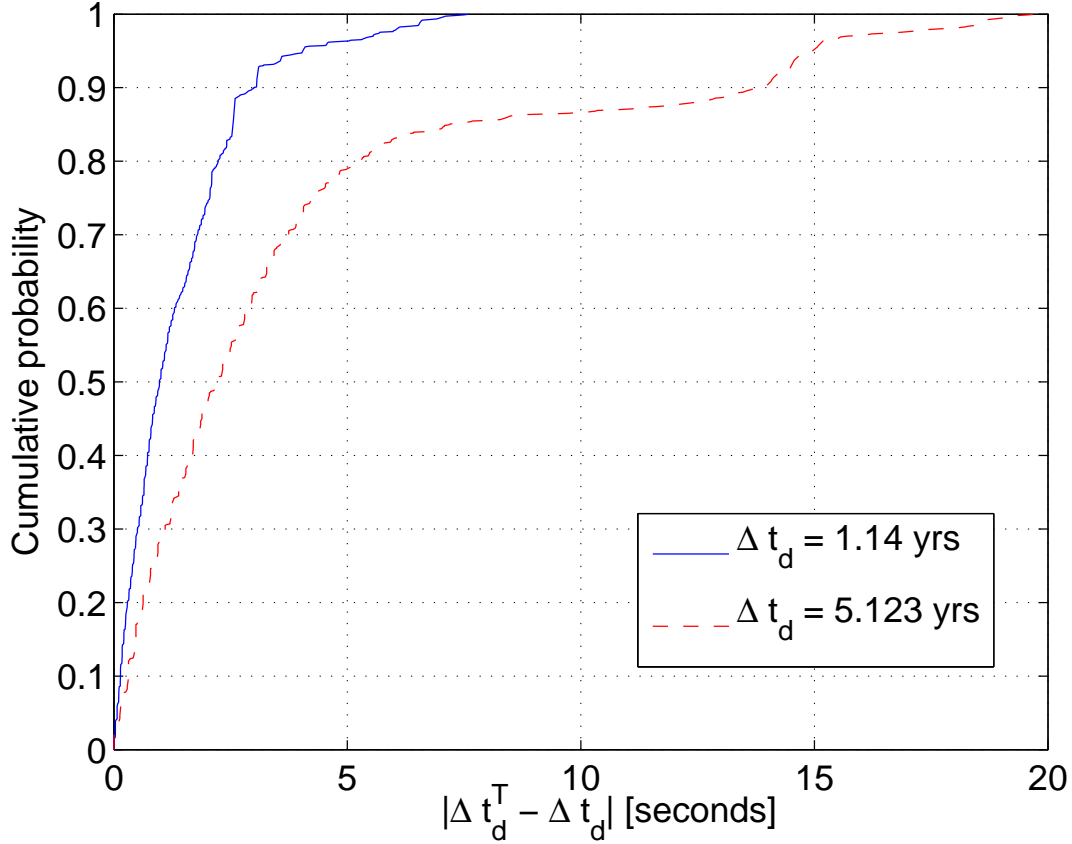}
\label{fig:ecdf_td}
\end{figure} 
\end{center}

Now what are the effect of the errors in the ${\bf p}^T$ onto estimates
of $\Gamma$ and $\Delta t_d$? 
Fig. \ref{fig:ecdf_Gamma} and Fig. \ref{fig:ecdf_td} show 
the cumulative probabilities of the absolute errors in $|\Gamma^T - \Gamma|$ 
and $|\Delta t_d^T - \Delta t_d|$.  From Fig.\ref{fig:ecdf_Gamma}, we conclude that
errors in ${\bf p}^T$ cause errors of less than $2.5\times 10^{-8}$ in $\Gamma^T$ ($\Gamma =
4.9\times 10^{-8}$) for
$35\%$ of the simulation for $\Delta t_d = 1.14$ years and  
$70\%$ of the simulation for $\Delta t_d = 5.1$ years.

\begin{center}
\begin{figure}
\caption{
A scatter plot of maximum locations of $\zeta$ in the
 $M_{cz}^T/M_{cz} - 1$  and $t_{c}^T/t_{c} - 1$ 
plane for 1100 sets of randomly chosen source,
 lens, and detector parameters. Here $\Delta t_d = 1.14$ years. 
Symbols indicate $|\Gamma^T - \Gamma|$ values: 
dots for $0 \le |\Gamma^T - \Gamma| < 3.33\times 10^{-8}$, 
crosses for $3.33\times10^{-8} \le |\Gamma^T - \Gamma| < 6.67\times 10^{-8}$, and 
squares for $6.67\times10^{-8} \le |\Gamma^T - \Gamma|$. The true value
 of  $\Gamma$ is $4.9\times 10^{-8}$.
}
\centering\includegraphics[width=16cm]{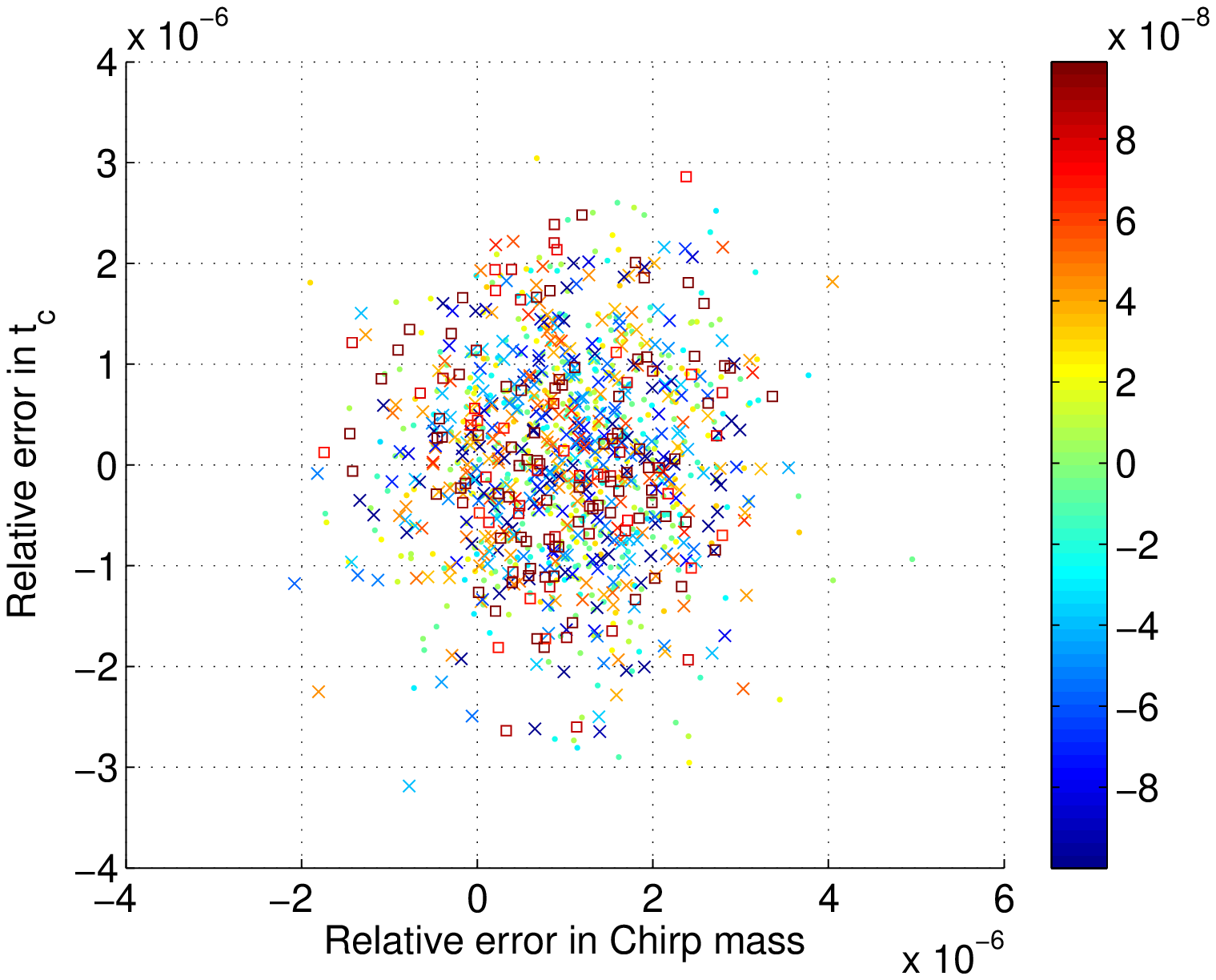}
\label{fig:mass_tc_gamma_map_largeError}
\end{figure} 
\end{center}

Finally, the scatter plot Fig. \ref{fig:mass_tc_gamma_map_largeError} shows
how the errors in $M_{cz}$ and $t_c$ affect our estimate of
$\Gamma$. Since $M_{cz,1}/M_{cz,2} = \Gamma$, we can determine $\Gamma$
using estimates of $M_{cz,j}$, if this ratio can be determined accurately enough. 
This figure shows that even though we admit errors in $M_{cz}$ of
order $10^{-6}$ (note that the injected signals have $\Gamma =
4.9\times10^{-8}$), $\Gamma$ can be determined at less than $10^{-7}$
accuracy in our simulation, using our filtered
cross-correlation method.

\section{Precision to which we could measure $\theta\beta_{\perp}$}
\label{sec:accuracy}

In reality, there is a noise in detector's outputs, so the Eqs. 
(\ref{eq:detecter_output_without_noise}) and
(\ref{eq:detecter_output_cross_correlation_without_noise}) should be 
actually 
\begin{eqnarray}
s_A(t) &=& \sum_{j=1,2} h_j(t) + n_A(t) \,\,\, {\rm for} \,\,\, t_{e,1}-\Delta t_d \le t \le t_{e,1}, \\
s_B(t) &=& h_{j=2}(t) + n_B(t) \,\,\, {\rm for} \,\,\, t_{e,1} \le t \le t_{e,2}, \\
\end{eqnarray}
where $n_{k=A,B}(t)$ are the detector's noise.
The presence of noise affects firstly the estimates of the filter
parameters ${\bf p}^T$ when one use matched filtering to estimate
those. Secondly, it affects the estimates of $\Gamma^T$ and $\Delta t_d^T$
when we use a filtered cross-correlation technique. We studied the
former effect in the previous section, and here we study the combination
of them. In the presence of the noise, $\zeta$ statistic should be 
\begin{eqnarray}
\zeta &=& 
4 Re \int_{f_i}^{f_e}
\frac{s_{A}(f) s_{B}^*(f)}{S_h(f)}\cos(  2\pi f \Delta \tilde t^T_{d,1}
+ \Theta(f,\mathbf p^T)) df  
\nonumber \\
&=& 
\zeta_0 + 
4 
Re \sum_{\alpha}\int_{f_i}^{f_e}
\frac{n_A(f)h_{\alpha,2}^{L*}(f)}{S_h(f)}\cos(  2\pi f \Delta \tilde t^T_{d,1}
+ \Theta(f,\mathbf p^T)) df
\nonumber \\
&+& 
4 
Re\sum_{\alpha} \int_{f_i}^{f_e}
\frac{n_B(f)h_{\alpha,1}^{L*}(f)}{S_h(f)}\cos(  2\pi f \Delta \tilde t^T_{d,1}
+ \Theta(f,\mathbf p^T)) df
\nonumber \\
&+& 
4 
Re \int_{f_i}^{f_e}
\frac{n_{A}(f)n_{B}^{*}(f)}{S_h(f)}\cos(  2\pi f \Delta \tilde t^T_{d,1}
+ \Theta(f,\mathbf p^T)) df.
\label{eq:zeta_with_noise}
\end{eqnarray} 
$\zeta_0$ is $\zeta$ statistic without noise given by Eq. (\ref{eq:zeta_statistic}).
As is in the previous section, 
we first generate a wave with a randomly chosen source parameters set
${\bar \theta_L,\bar \phi_L,\bar \theta_S,\bar \phi_S,\phi_c,\bar
\phi_0,\alpha_0}$ where other parameters are taken from the table \ref{tbl:Q0957+561}. 
We then assume template parameters ${\bf p}^T$ by adding Gaussian errors to 
the true parameters of the generated signal. 
The standard deviations of the errors are the same as before. 
For each realization of the wave, we further add random Gaussian noise $n_j(f)$
satisfying $<n(f)n^*(f')> = S_h(f)\delta(f-f')/2$ where $<...>$ denotes ensemble
average. We then 
compute our statistic $\zeta$ and find $\Gamma$ and $\Delta t_d$ that
maximize $|\zeta|$. For each generation of a wave, noise frequency
series is generated 1000 times. We went through these steps 10 times, so 
that
we obtained total of $10 \times 1000$ estimates of $\Gamma$ and $\Delta t_d$. 

\begin{center}
\begin{figure}
\caption{
Cumulative probability distributions of $|\Gamma^T - \Mean(\Gamma^T)|$ 
for  8 waves with different source-lens-detector parameters sets   in the presence of noise.     
Here $\Delta t_d = 1.41$ years. 
See Sec. \ref{sec:accuracy} for details. 
}
\centering\includegraphics[width=16cm]{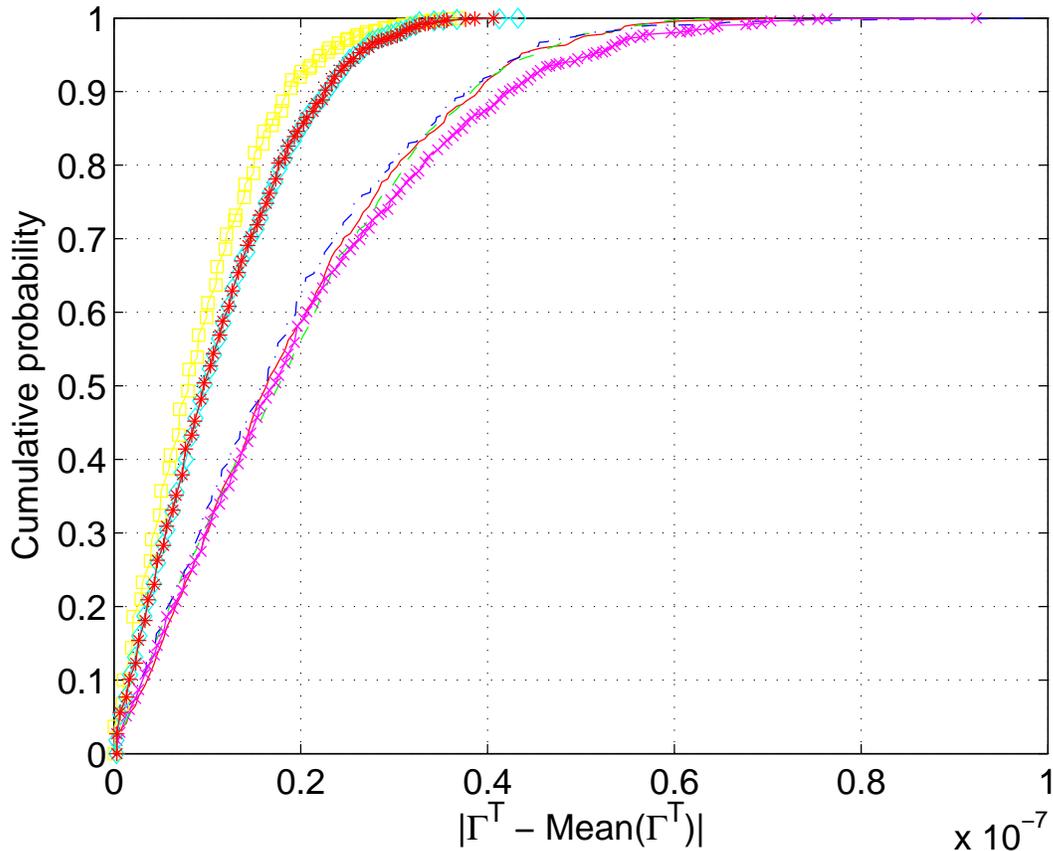}
\label{fig:ecdf_Gamma_with_noise_fmin0113_mc10x1000}
\end{figure} 
\end{center}

\begin{center}
\begin{figure}
\caption{
Cumulative probability distributions of $|\Gamma^T - \Mean(\Gamma^T)|$ 
for 10 waves with different source-lens-detector parameters sets  in the presence of noise.     
Here $\Delta t_d = 5.1$ years. 
See Sec. \ref{sec:accuracy} for details. 
}
\centering\includegraphics[width=16cm]{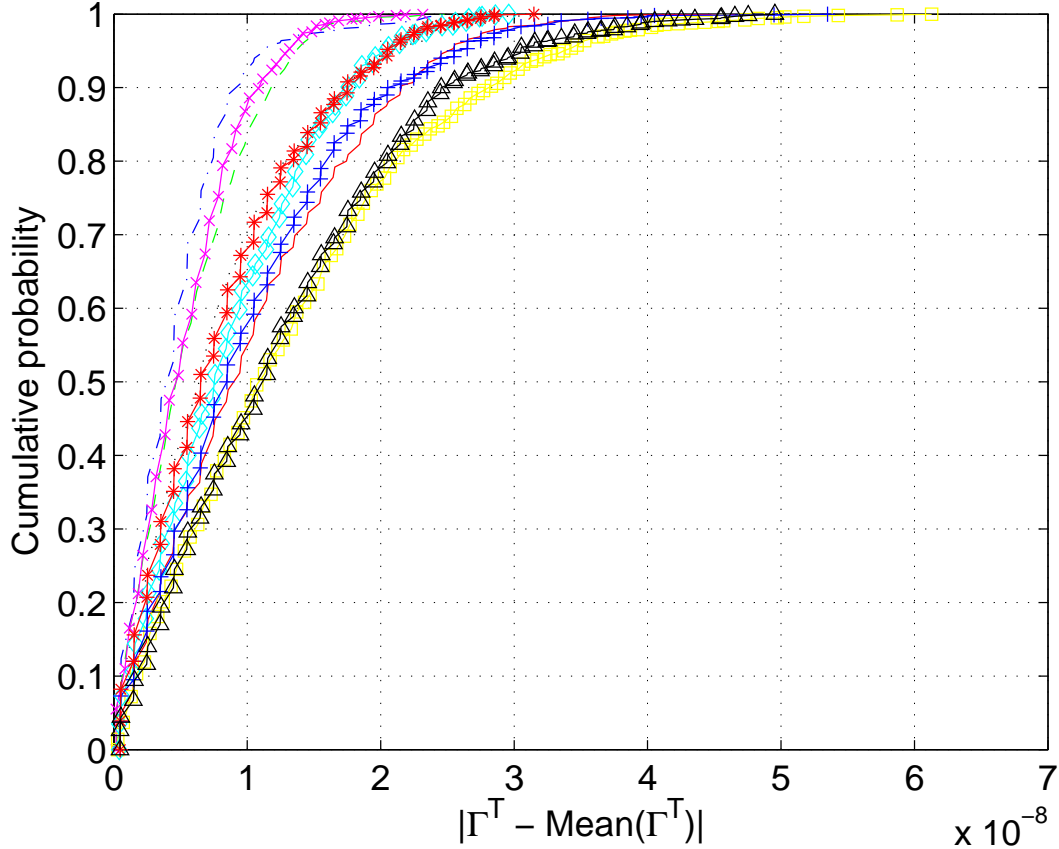}
\label{fig:ecdf_Gamma_with_noise_fmin0065_mc10x1000}
\end{figure} 
\end{center}

The results of the simulation are as follows. 
The cumulative probabilities of the errors in $|\Gamma^T - \Mean(\Gamma^T)|$ are
shown for 8 out of 10 waves in
Fig. \ref{fig:ecdf_Gamma_with_noise_fmin0113_mc10x1000} (For the excluded 2
waves the maximum of the $\zeta_0$ appears to close to the search region
boundary and we could not compute the cumulative probabilities
properly). 
Here $\Mean(\Gamma^T)$ for each source parameters set is the mean of $\Gamma^T$ over 1000 realizations
of the noise series and does not necessarily equals the true value of
$\Gamma$, as studied in the previous section. 
The standard deviation of the errors is $1.9\times 10^{-8}$ and $90\%$
of the time the error is less than $3\times 10^{-8}$ on average of 8 waves.
We also performed the same analysis but with $\Delta t_d = 1.14$ years
replaced  
by $\Delta t_d = 5.1$ years (so that $f_i = 0.113$ Hz by $f_i = 0.065$
Hz). Fig. \ref{fig:ecdf_Gamma_with_noise_fmin0065_mc10x1000} shows the
result for this longer time delay (and thus longer integration time of 
5 years).
The standard deviation of the errors in this case is $1.1\times 10^{-8}$ and $90\%$
of the time the error is less than $2\times 10^{-8}$ on average of 10 waves.

Finally, we performed the same analysis for $v = 1000 \kms$ or $\Gamma =
10^{-7}$, and obtained quantitatively the same results as above. 
Combining the results in the previous sections, we conclude
that $\Gamma^T$ would be determined, in our example,  with the future
(one-triangle) BBO/DECIGO detectors 
\begin{eqnarray}
\Gamma^T = \Gamma 
\pm 5.5\times 10^{-8}
\pm 1.9\times 10^{-8}
 \end{eqnarray} 
for 1.14 years integration and 
\begin{eqnarray}
\Gamma^T = \Gamma
\pm 2.0\times 10^{-8}
\pm 1.1\times 10^{-8}
 \end{eqnarray} 
for 5.1 years integration, where the first error in each equation is due to the
errors in the estimates of the filter parameters ${\bf p}$  and
the second error comes from the noise terms in the filtered
cross-correlation analysis in Eq. (\ref{eq:zeta_with_noise}).

\section{Summary}
\label{sec:summary}

In this paper, we considered a gravitational lensing phenomena of 
gravitational waves in the case  
when the geometrical optics approximation applies.  
To be specific, the source was assumed to be 
a coalescing neutron stars binary at a cosmological distance and 
a detector to be the planned space-borne detector (one-triangle)
BBO/DECIGO.  We then proposed a filtered cross-correlation method of extracting the relative
transverse velocity of the source-lens-observer system using 
an interference term.

We performed series of simulations to study correlations among
parameters and to estimate errors due to detector's noise.   
In these simulations, we adopted lens parameters taken from the lens system Q0957+561
\cite{Walsh1979Natur,CASTLE} as a reference for a demonstration purpose, in which the time delay is
$\Delta t_d = 1.41$ years. 
With our method, we found the absolute error in $\Gamma = \theta\beta_{\perp}$ 
($\theta$ is the images separation and $\beta_{\perp}$ is the relative
transverse velocity),  with the future
(one-triangle) BBO/DECIGO detectors to be 
$
\delta (\Gamma) =  
\pm 5.5\times 10^{-8}
\pm 1.9\times 10^{-8}
$
for the time delay of 1.14 years, where the first error in the equation is due to the
error in the estimates of the filter parameters ${\bf p}$  and
the second error comes from the noise terms in the filtered
cross-correlation analysis in Eq. (\ref{eq:zeta_with_noise}).  
The errors in the equation are the standard deviation estimated
in our simulations. 
We also
performed the same analysis but with $\Delta t_d = 5.1$ years. In this
case, we found 
$
\delta (\Gamma) =  
\pm 2.0\times 10^{-8}
\pm 1.1\times 10^{-8}
$.

Although the
probabilities of lens phenomena is highly uncertain mainly due to the 
fact that the event rate of neutron stars binaries coalescence is
highly uncertain, one observation of transverse velocity of 
a neutron stars binary may give us valuable information of such population.

\begin{acknowledgments}
We would like to thank Ryuichi Takahashi of 
Hirosaki University, Japan and 
Nobuhiro Okabe of Academia Sinica, Institute of 
Astronomy and Astrophysics, in Taiwan for useful discussion.
\end{acknowledgments}


\begin{thebibliography}{30}
\expandafter\ifx\csname natexlab\endcsname\relax\def\natexlab#1{#1}\fi
\expandafter\ifx\csname bibnamefont\endcsname\relax
  \def\bibnamefont#1{#1}\fi
\expandafter\ifx\csname bibfnamefont\endcsname\relax
  \def\bibfnamefont#1{#1}\fi
\expandafter\ifx\csname citenamefont\endcsname\relax
  \def\citenamefont#1{#1}\fi
\expandafter\ifx\csname url\endcsname\relax
  \def\url#1{\texttt{#1}}\fi
\expandafter\ifx\csname urlprefix\endcsname\relax\def\urlprefix{URL }\fi
\providecommand{\bibinfo}[2]{#2}
\providecommand{\eprint}[2][]{\url{#2}}

\bibitem[{\citenamefont{{Thompson} et~al.}(2001)\citenamefont{{Thompson},
  {Moran}, and {Swenson}}}]{ThompsonMoranSwensonBook}
\bibinfo{author}{\bibfnamefont{A.~R.} \bibnamefont{{Thompson}}},
  \bibinfo{author}{\bibfnamefont{J.~M.} \bibnamefont{{Moran}}},
  \bibnamefont{and} \bibinfo{author}{\bibfnamefont{G.~W.}
  \bibnamefont{{Swenson}}, \bibfnamefont{Jr.}},
  \emph{\bibinfo{title}{{Interferometry and Synthesis in Radio Astronomy, 2nd
  Edition}}} (\bibinfo{publisher}{New York : Wiley}, \bibinfo{year}{2001}).

\bibitem[{\citenamefont{{Frail} et~al.}(1997)\citenamefont{{Frail}, {Kulkarni},
  {Nicastro}, {Feroci}, and {Taylor}}}]{Frail1997}
\bibinfo{author}{\bibfnamefont{D.~A.} \bibnamefont{{Frail}}},
  \bibinfo{author}{\bibfnamefont{S.~R.} \bibnamefont{{Kulkarni}}},
  \bibinfo{author}{\bibfnamefont{L.}~\bibnamefont{{Nicastro}}},
  \bibinfo{author}{\bibfnamefont{M.}~\bibnamefont{{Feroci}}}, \bibnamefont{and}
  \bibinfo{author}{\bibfnamefont{G.~B.} \bibnamefont{{Taylor}}},
  \bibinfo{journal}{\nat} \textbf{\bibinfo{volume}{389}}, \bibinfo{pages}{261}
  (\bibinfo{year}{1997}).

\bibitem[{\citenamefont{Cutler and Thorne}(2002)}]{Cutler2002}
\bibinfo{author}{\bibfnamefont{C.}~\bibnamefont{Cutler}} \bibnamefont{and}
  \bibinfo{author}{\bibfnamefont{K.~S.} \bibnamefont{Thorne}},
  \emph{\bibinfo{title}{{An overview of gravitational-wave sources}}}
  (\bibinfo{year}{2002}), \eprint{gr-qc/0204090}.

\bibitem[{\citenamefont{{Abramovici} et~al.}(1992)\citenamefont{{Abramovici},
  {Althouse}, {Drever}, {Gursel}, {Kawamura}, {Raab}, {Shoemaker}, {Sievers},
  {Spero}, and {Thorne}}}]{LIGO1992}
\bibinfo{author}{\bibfnamefont{A.}~\bibnamefont{{Abramovici}}},
  \bibinfo{author}{\bibfnamefont{W.~E.} \bibnamefont{{Althouse}}},
  \bibinfo{author}{\bibfnamefont{R.~W.~P.} \bibnamefont{{Drever}}},
  \bibinfo{author}{\bibfnamefont{Y.}~\bibnamefont{{Gursel}}},
  \bibinfo{author}{\bibfnamefont{S.}~\bibnamefont{{Kawamura}}},
  \bibinfo{author}{\bibfnamefont{F.~J.} \bibnamefont{{Raab}}},
  \bibinfo{author}{\bibfnamefont{D.}~\bibnamefont{{Shoemaker}}},
  \bibinfo{author}{\bibfnamefont{L.}~\bibnamefont{{Sievers}}},
  \bibinfo{author}{\bibfnamefont{R.~E.} \bibnamefont{{Spero}}},
  \bibnamefont{and} \bibinfo{author}{\bibfnamefont{K.~S.}
  \bibnamefont{{Thorne}}}, \bibinfo{journal}{Science}
  \textbf{\bibinfo{volume}{256}}, \bibinfo{pages}{325} (\bibinfo{year}{1992}).

\bibitem[{\citenamefont{{Kuroda} et~al.}(1999)\citenamefont{{Kuroda}, {Ohashi},
  {Miyoki}, {Tatsumi}, {Sato}, {Ishizuka}, {Fujimoto}, {Kawamura}, {Takahashi},
  {Yamazaki} et~al.}}]{LCGT1999}
\bibinfo{author}{\bibfnamefont{K.}~\bibnamefont{{Kuroda}}},
  \bibinfo{author}{\bibfnamefont{M.}~\bibnamefont{{Ohashi}}},
  \bibinfo{author}{\bibfnamefont{S.}~\bibnamefont{{Miyoki}}},
  \bibinfo{author}{\bibfnamefont{D.}~\bibnamefont{{Tatsumi}}},
  \bibinfo{author}{\bibfnamefont{S.}~\bibnamefont{{Sato}}},
  \bibinfo{author}{\bibfnamefont{H.}~\bibnamefont{{Ishizuka}}},
  \bibinfo{author}{\bibfnamefont{M.-K.} \bibnamefont{{Fujimoto}}},
  \bibinfo{author}{\bibfnamefont{S.}~\bibnamefont{{Kawamura}}},
  \bibinfo{author}{\bibfnamefont{R.}~\bibnamefont{{Takahashi}}},
  \bibinfo{author}{\bibfnamefont{T.}~\bibnamefont{{Yamazaki}}},
  \bibnamefont{et~al.}, \bibinfo{journal}{International Journal of Modern
  Physics D} \textbf{\bibinfo{volume}{8}}, \bibinfo{pages}{557}
  (\bibinfo{year}{1999}).

\bibitem[{\citenamefont{Danzmann and {et al.}}(1996)}]{LISA1996}
\bibinfo{author}{\bibfnamefont{K.}~\bibnamefont{Danzmann}} \bibnamefont{and}
  \bibinfo{author}{\bibnamefont{{et al.}}}, \emph{\bibinfo{title}{{LISA
  Pre-Phase A Report}}} (\bibinfo{year}{1996}),
  \bibinfo{note}{{Max-Planck-Institut fur Quantenoptik, Report No. MPQ 208,
  Garching, Germany}}.

\bibitem[{\citenamefont{Phinney and {et al.}}(2003)}]{Phinney2003}
\bibinfo{author}{\bibfnamefont{E.~S.} \bibnamefont{Phinney}} \bibnamefont{and}
  \bibinfo{author}{\bibnamefont{{et al.}}}, \emph{\bibinfo{title}{{The Big Bang
  Observer, NASA Mission Concept Study}}} (\bibinfo{year}{2003}).

\bibitem[{\citenamefont{{Seto} et~al.}(2001)\citenamefont{{Seto}, {Kawamura},
  and {Nakamura}}}]{SetoKawamuraNakamura2001}
\bibinfo{author}{\bibfnamefont{N.}~\bibnamefont{{Seto}}},
  \bibinfo{author}{\bibfnamefont{S.}~\bibnamefont{{Kawamura}}},
  \bibnamefont{and}
  \bibinfo{author}{\bibfnamefont{T.}~\bibnamefont{{Nakamura}}},
  \bibinfo{journal}{Physical Review Letters} \textbf{\bibinfo{volume}{87}},
  \bibinfo{pages}{221103} (\bibinfo{year}{2001}).

\bibitem[{\citenamefont{Kochanek et~al.}(2006)\citenamefont{Kochanek, Falco,
  Impey, Lehar, McLeod, and Rix}}]{CASTLE}
\bibinfo{author}{\bibfnamefont{C.}~\bibnamefont{Kochanek}},
  \bibinfo{author}{\bibfnamefont{E.}~\bibnamefont{Falco}},
  \bibinfo{author}{\bibfnamefont{C.}~\bibnamefont{Impey}},
  \bibinfo{author}{\bibfnamefont{J.}~\bibnamefont{Lehar}},
  \bibinfo{author}{\bibfnamefont{B.}~\bibnamefont{McLeod}}, \bibnamefont{and}
  \bibinfo{author}{\bibfnamefont{H.-W.} \bibnamefont{Rix}},
  \emph{\bibinfo{title}{{CfA-Arizona Space Telescope LEns Survey of
  gravitational lenses}}} (\bibinfo{year}{2006}),
  \bibinfo{note}{http://cfa-www.harvard.edu/castles/}.

\bibitem[{\citenamefont{{Mu{\~n}oz} et~al.}(1998)\citenamefont{{Mu{\~n}oz},
  {Falco}, {Kochanek}, {Leh{\'a}r}, {McLeod}, {Impey}, {Rix}, and
  {Peng}}}]{CASTLE1998}
\bibinfo{author}{\bibfnamefont{J.~A.} \bibnamefont{{Mu{\~n}oz}}},
  \bibinfo{author}{\bibfnamefont{E.~E.} \bibnamefont{{Falco}}},
  \bibinfo{author}{\bibfnamefont{C.~S.} \bibnamefont{{Kochanek}}},
  \bibinfo{author}{\bibfnamefont{J.}~\bibnamefont{{Leh{\'a}r}}},
  \bibinfo{author}{\bibfnamefont{B.~A.} \bibnamefont{{McLeod}}},
  \bibinfo{author}{\bibfnamefont{C.~D.} \bibnamefont{{Impey}}},
  \bibinfo{author}{\bibfnamefont{H.-W.} \bibnamefont{{Rix}}}, \bibnamefont{and}
  \bibinfo{author}{\bibfnamefont{C.~Y.} \bibnamefont{{Peng}}},
  \bibinfo{journal}{Astrophysics and Space Science}
  \textbf{\bibinfo{volume}{263}}, \bibinfo{pages}{51} (\bibinfo{year}{1998}).

\bibitem[{\citenamefont{{Peterson} and {Falk}}(1991)}]{Peterson1991}
\bibinfo{author}{\bibfnamefont{J.~B.} \bibnamefont{{Peterson}}}
  \bibnamefont{and} \bibinfo{author}{\bibfnamefont{T.}~\bibnamefont{{Falk}}},
  \bibinfo{journal}{\apjl} \textbf{\bibinfo{volume}{374}}, \bibinfo{pages}{L5}
  (\bibinfo{year}{1991}).

\bibitem[{\citenamefont{{Takahashi}}(2004)}]{Takahashi2004}
\bibinfo{author}{\bibfnamefont{R.}~\bibnamefont{{Takahashi}}},
  \bibinfo{journal}{Astronomy and Astrophysics} \textbf{\bibinfo{volume}{423}},
  \bibinfo{pages}{787} (\bibinfo{year}{2004}).

\bibitem[{\citenamefont{{Matsunaga} and {Yamamoto}}(2006)}]{Matsunaga2006}
\bibinfo{author}{\bibfnamefont{N.}~\bibnamefont{{Matsunaga}}} \bibnamefont{and}
  \bibinfo{author}{\bibfnamefont{K.}~\bibnamefont{{Yamamoto}}},
  \bibinfo{journal}{Journal of Cosmology and Astro-Particle Physics}
  \textbf{\bibinfo{volume}{1}}, \bibinfo{pages}{23} (\bibinfo{year}{2006}).

\bibitem[{\citenamefont{{Takahashi}}(2006)}]{Takahashi2006}
\bibinfo{author}{\bibfnamefont{R.}~\bibnamefont{{Takahashi}}},
  \bibinfo{journal}{\apj} \textbf{\bibinfo{volume}{644}}, \bibinfo{pages}{80}
  (\bibinfo{year}{2006}).

\bibitem[{\citenamefont{{Yoo} et~al.}(2007)\citenamefont{{Yoo}, {Nakao},
  {Kozaki}, and {Takahashi}}}]{Yoo2007}
\bibinfo{author}{\bibfnamefont{C.-M.} \bibnamefont{{Yoo}}},
  \bibinfo{author}{\bibfnamefont{K.-i.} \bibnamefont{{Nakao}}},
  \bibinfo{author}{\bibfnamefont{H.}~\bibnamefont{{Kozaki}}}, \bibnamefont{and}
  \bibinfo{author}{\bibfnamefont{R.}~\bibnamefont{{Takahashi}}},
  \bibinfo{journal}{\apj} \textbf{\bibinfo{volume}{655}}, \bibinfo{pages}{691}
  (\bibinfo{year}{2007}).

\bibitem[{\citenamefont{{Schneider} et~al.}(1992)\citenamefont{{Schneider},
  {Ehlers}, and {Falco}}}]{Schneider1992}
\bibinfo{author}{\bibfnamefont{P.}~\bibnamefont{{Schneider}}},
  \bibinfo{author}{\bibfnamefont{J.}~\bibnamefont{{Ehlers}}}, \bibnamefont{and}
  \bibinfo{author}{\bibfnamefont{E.~E.} \bibnamefont{{Falco}}},
  \emph{\bibinfo{title}{{Gravitational Lenses}}}
  (\bibinfo{publisher}{Springer-Verlag Berlin Heidelberg New York},
  \bibinfo{year}{1992}).

\bibitem[{\citenamefont{{Baraldo} et~al.}(1999)\citenamefont{{Baraldo},
  {Hosoya}, and {Nakamura}}}]{Baraldo1999}
\bibinfo{author}{\bibfnamefont{C.}~\bibnamefont{{Baraldo}}},
  \bibinfo{author}{\bibfnamefont{A.}~\bibnamefont{{Hosoya}}}, \bibnamefont{and}
  \bibinfo{author}{\bibfnamefont{T.~T.} \bibnamefont{{Nakamura}}},
  \bibinfo{journal}{\prd} \textbf{\bibinfo{volume}{59}},
  \bibinfo{pages}{083001} (\bibinfo{year}{1999}).

\bibitem[{\citenamefont{{Birkinshaw} and {Gull}}(1983)}]{Birkinshaw1983}
\bibinfo{author}{\bibfnamefont{M.}~\bibnamefont{{Birkinshaw}}}
  \bibnamefont{and} \bibinfo{author}{\bibfnamefont{S.~F.}
  \bibnamefont{{Gull}}}, \bibinfo{journal}{\nat}
  \textbf{\bibinfo{volume}{302}}, \bibinfo{pages}{315} (\bibinfo{year}{1983}).

\bibitem[{\citenamefont{{Pyne} and {Birkinshaw}}(1993)}]{Pyne1993}
\bibinfo{author}{\bibfnamefont{T.}~\bibnamefont{{Pyne}}} \bibnamefont{and}
  \bibinfo{author}{\bibfnamefont{M.}~\bibnamefont{{Birkinshaw}}},
  \bibinfo{journal}{\apj} \textbf{\bibinfo{volume}{415}}, \bibinfo{pages}{459}
  (\bibinfo{year}{1993}).

\bibitem[{\citenamefont{{Aso} et~al.}(2002)\citenamefont{{Aso}, {Hattori}, and
  {Futamase}}}]{Aso2002}
\bibinfo{author}{\bibfnamefont{O.}~\bibnamefont{{Aso}}},
  \bibinfo{author}{\bibfnamefont{M.}~\bibnamefont{{Hattori}}},
  \bibnamefont{and}
  \bibinfo{author}{\bibfnamefont{T.}~\bibnamefont{{Futamase}}},
  \bibinfo{journal}{\apjl} \textbf{\bibinfo{volume}{576}}, \bibinfo{pages}{L5}
  (\bibinfo{year}{2002}).

\bibitem[{\citenamefont{{Grieger} et~al.}(1986)\citenamefont{{Grieger},
  {Kayser}, and {Refsdal}}}]{Grieger1986}
\bibinfo{author}{\bibfnamefont{B.}~\bibnamefont{{Grieger}}},
  \bibinfo{author}{\bibfnamefont{R.}~\bibnamefont{{Kayser}}}, \bibnamefont{and}
  \bibinfo{author}{\bibfnamefont{S.}~\bibnamefont{{Refsdal}}},
  \bibinfo{journal}{\nat} \textbf{\bibinfo{volume}{324}}, \bibinfo{pages}{126}
  (\bibinfo{year}{1986}).

\bibitem[{\citenamefont{{Gould}}(1995)}]{Gould1995}
\bibinfo{author}{\bibfnamefont{A.}~\bibnamefont{{Gould}}},
  \bibinfo{journal}{\apj} \textbf{\bibinfo{volume}{444}}, \bibinfo{pages}{556}
  (\bibinfo{year}{1995}).

\bibitem[{\citenamefont{{Molnar} and {Birkinshaw}}(2003)}]{Molnar2003}
\bibinfo{author}{\bibfnamefont{S.~M.} \bibnamefont{{Molnar}}} \bibnamefont{and}
  \bibinfo{author}{\bibfnamefont{M.}~\bibnamefont{{Birkinshaw}}},
  \bibinfo{journal}{\apj} \textbf{\bibinfo{volume}{586}}, \bibinfo{pages}{731}
  (\bibinfo{year}{2003}).

\bibitem[{\citenamefont{{Takahashi} and
  {Nakamura}}(2003)}]{TakahashiNakamura2003}
\bibinfo{author}{\bibfnamefont{R.}~\bibnamefont{{Takahashi}}} \bibnamefont{and}
  \bibinfo{author}{\bibfnamefont{T.}~\bibnamefont{{Nakamura}}},
  \bibinfo{journal}{\apj} \textbf{\bibinfo{volume}{595}}, \bibinfo{pages}{1039}
  (\bibinfo{year}{2003}).

\bibitem[{\citenamefont{{Cutler}}(1998)}]{Cutler1998}
\bibinfo{author}{\bibfnamefont{C.}~\bibnamefont{{Cutler}}},
  \bibinfo{journal}{\prd} \textbf{\bibinfo{volume}{57}}, \bibinfo{pages}{7089}
  (\bibinfo{year}{1998}).

\bibitem[{\citenamefont{{Cutler} and {Harms}}(2006)}]{CutlerHarms2003}
\bibinfo{author}{\bibfnamefont{C.}~\bibnamefont{{Cutler}}} \bibnamefont{and}
  \bibinfo{author}{\bibfnamefont{J.}~\bibnamefont{{Harms}}},
  \bibinfo{journal}{\prd} \textbf{\bibinfo{volume}{73}},
  \bibinfo{pages}{042001} (\bibinfo{year}{2006}).

\bibitem[{\citenamefont{{Kayser} et~al.}(1986)\citenamefont{{Kayser},
  {Refsdal}, and {Stabell}}}]{Kayser1986}
\bibinfo{author}{\bibfnamefont{R.}~\bibnamefont{{Kayser}}},
  \bibinfo{author}{\bibfnamefont{S.}~\bibnamefont{{Refsdal}}},
  \bibnamefont{and}
  \bibinfo{author}{\bibfnamefont{R.}~\bibnamefont{{Stabell}}},
  \bibinfo{journal}{\aap} \textbf{\bibinfo{volume}{166}}, \bibinfo{pages}{36}
  (\bibinfo{year}{1986}).

\bibitem[{\citenamefont{{Wucknitz} and {Sperhake}}(2004)}]{Wicknitz2004}
\bibinfo{author}{\bibfnamefont{O.}~\bibnamefont{{Wucknitz}}} \bibnamefont{and}
  \bibinfo{author}{\bibfnamefont{U.}~\bibnamefont{{Sperhake}}},
  \bibinfo{journal}{\prd} \textbf{\bibinfo{volume}{69}},
  \bibinfo{pages}{063001} (\bibinfo{year}{2004}).

\bibitem[{\citenamefont{{Walsh} et~al.}(1979)\citenamefont{{Walsh}, {Carswell},
  and {Weymann}}}]{Walsh1979Natur}
\bibinfo{author}{\bibfnamefont{D.}~\bibnamefont{{Walsh}}},
  \bibinfo{author}{\bibfnamefont{R.~F.} \bibnamefont{{Carswell}}},
  \bibnamefont{and} \bibinfo{author}{\bibfnamefont{R.~J.}
  \bibnamefont{{Weymann}}}, \bibinfo{journal}{\nat}
  \textbf{\bibinfo{volume}{279}}, \bibinfo{pages}{381} (\bibinfo{year}{1979}).

\bibitem[{\citenamefont{{Crowder} and {Cornish}}(2005)}]{Crowder2005}
\bibinfo{author}{\bibfnamefont{J.}~\bibnamefont{{Crowder}}} \bibnamefont{and}
  \bibinfo{author}{\bibfnamefont{N.~J.} \bibnamefont{{Cornish}}},
  \bibinfo{journal}{\prd} \textbf{\bibinfo{volume}{72}},
  \bibinfo{pages}{083005} (\bibinfo{year}{2005}).

\end{thebibliography}
\end{document}